\documentclass[preprint2]{aastex}
\newcommand{\magn}[1]{\mbox{$\rm #1^m$}} 
\newcommand{\magpt}[2]{\mbox{$\rm #1\hspace{-0.25em}\stackrel{m}{.}
      \hspace{-1.0mm}#2$}}                             
\newcommand\bmin{\hbox{$.\!\!{'}$}}

\newcommand\ebv{$E_{B-V}$}
\newcommand\teff{$ T_{\rm eff}$}

\newcommand\logg{$\log g$}

\newcommand\logLL{$\log{\frac{L}{L_\odot}}$}

\newcommand{\Msolar}{\mbox{\,$\rm M_{\odot}$}} 

\begin{document}
\title{Hot Stars in Globular Clusters -- A Spectroscopist's View\\}
\author{S. Moehler\\}
\affil{Dr. Remeis-Sternwarte, Astronomisches Institut der Universit\"at 
Erlangen-N\"urnberg, Sternwartstr. 7, D-96049 Bamberg, Germany
(e-mail: ai13@sternwarte.uni-erlangen.de}

\begin{abstract}
Globular clusters are ideal laboratories to study the evolution of low-mass 
stars. In this work we concentrate on three types of hot stars observed in 
globular clusters: horizontal branch stars, UV bright stars, and white dwarfs.
After providing some historical background and information on gaps and blue tails 
we discuss extensively
hot horizontal branch stars in metal-poor globular clusters, esp.
their abundance anomalies and the consequences for the determination of their 
atmospheric parameters and evolutionary status. Hot horizontal branch 
stars in metal-rich globular clusters are found to form a small, but rather 
inhomogeneous group that cannot be explained by {\em one} evolutionary 
scenario. 
Hot UV bright stars show a lack of classic post-AGB stars that 
may explain the lack of planetary nebulae in globular clusters. 
Finally we discuss first results of spectroscopic observations of white 
dwarfs in globular clusters.
\end{abstract}
\maketitle

\section{Historical Background\label{history}}

Globular clusters are the closest approximation to a physicist's laboratory
in astronomy. They are densely packed, gravitationally bound systems of
several thousands to about one million stars. The dimensions of the
globular clusters are small compared to their distance from us: half of the
light is generally emitted within a radius of less than 10~pc, whereas the
closest globular cluster has a distance of 2~kpc and 90\% lie more than
5~kpc away. We can thus safely assume that all stars within a globular
cluster lie at the same distance from us. With ages in the order of 
$10^{10}$~years globular clusters are among the oldest objects in our
Galaxy. Contrary to the field of the Galaxy globular clusters
formed stars only once in the beginning. Because the duration of that star
formation episode is short compared to the current age of the globular
clusters the stars within one globular cluster are essentially coeval. In
addition all stars within one globular cluster (with few exceptions)
show the same initial
abundance pattern (which may differ from one cluster to another). 

As we know today that Galactic globular clusters are old stellar systems 
people are often surprised by the presence of hot stars in these 
clusters since hot stars are usually associated with young stellar 
systems.  The following paragraphs will show that hot stars have been known to
exist in globular clusters for quite some time:

About a century ago
\citet{barn00} reported the detection of stars in globular clusters
that were much brighter on (blue-sensitive)
photographic plates than they appeared visually: 
{\it
``Of course the simple explanation of this peculiarity is that these stars, so
bright photographically and so faint visually, are shining with a much bluer
light than the stars which make up the main body of the clusters.''}

In 1915 Shapley started a project to obtain colours and magnitudes of
individual stars in globular and open clusters \citep{shap15a}. In the
first globular cluster \citep[M~3, ][]{shap15b} he found a double peaked
distribution of colours, with a red maximum and a blue secondary peak. He
noticed that -- in contrast to what was known for field dwarf (i.e.\ main
sequence) stars -- the stars in M~3 became bluer as they became fainter.
\citet[p.\ 130]{tebr27} used Shapley's data on M~3 and other clusters to
plot magnitude versus colour  (replacing luminosity and spectral type in
the Hertzsprung-Russell diagram) and thus produced the first
colour-magnitude diagrams\footnote{\citet[p.26, footnote]{shap30} disliked
the idea of plotting individual data points -- he thought that the small
number of measurements might lead to spurious results.} ({\sc
``Farbenhelligkeitsdiagramme''}). In these colour-magnitude diagrams
(CMD's) ten Bruggencate noted the presence of a red giant branch that
became bluer towards fainter magnitudes, in agreement with \citet{shap15b}.
 In addition, however, he saw a horizontal branch ({\sc ``horizontaler
Ast''}) that parted from the red giant branch and extended far to the blue
at constant brightness. \citet{gree39} observed a colour-magnitude diagram
for M~4 and noticed that -- while hot main-sequence stars were completely
missing -- there existed a group of bright stars above the horizontal
branch and on the blue side of the red giant branch. Similar stars appeared
also in the CMD's presented by \citet{arp55}. 

\begin{figure}[ht]
\vspace*{9.3cm}
\includegraphics{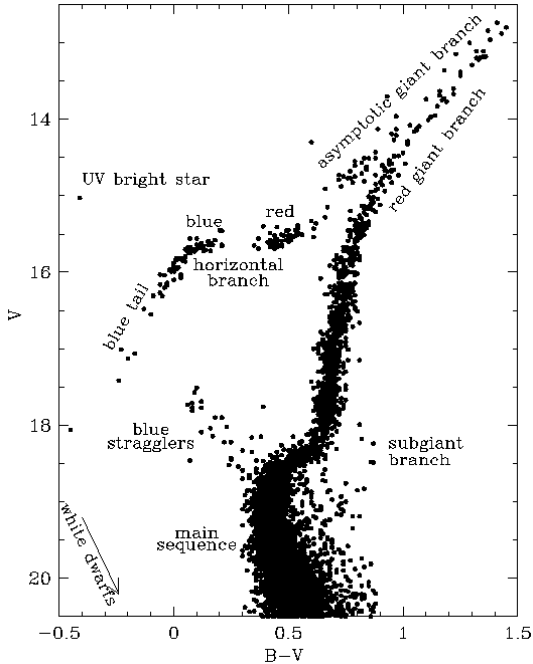}
\caption[Exemplary colour-magnitude diagram with the names of the principal 
sequences]
{Colour-magnitude diagram of M~3 \citep{buco94} with 
the names of the principal sequences.\label{ag_cmd}}
\end{figure}

As more CMD's of globular clusters were obtained it became obvious that the
horizontal branch morphology varied quite considerably between individual
clusters. The clusters observed by \citet{arp55} exhibited extensions of
the blue horizontal branch towards bluer colours and fainter visual
magnitudes, i.e. towards hotter temperatures\footnote{The change in slope
of the horizontal branch towards higher temperatures is caused by the
decreasing sensitivity of $B-V$ to temperature on one hand and by the
increasing bolometric correction for hotter stars (i.e. the maximum of
stellar flux is radiated at ever shorter wavelengths for increasing
temperatures, making stars fainter at $V$) on the other hand.} (see
Fig.~\ref{ag_cmd}). In some of Arp's CMD's (e.g. M~15, M~2) these {\bf blue
tails} show gaps at varying brightness
(see Sect.~\ref{sec-gaps} for details). 

About 25 years after their discovery first ideas about the nature of the
horizontal branch stars began to emerge: \citet{hosc55} 
were the first to identify the horizontal branch stars with 
post-red giant branch stars that burn helium in the central regions of
their cores.

\citet{sawa60}  noted a correlation between the metal
abundance and the horizontal branch morphology seen in globular cluster CMD's:
the horizontal branch (HB) became bluer with decreasing metallicity. 
\citet{faul66} managed for the first time to compute zero age horizontal
branch (ZAHB) models that qualitatively reproduced this trend of HB morphology
with metallicity without taking into account any mass loss but assuming
a rather high helium abundance of Y = 0.35. \citet{ibro70}, 
however, found that 
{\it
``In fact for the values of Y and Z most favored (Y $\ge$ 0.25 $\rightarrow$
0.28, Z = $10^{-3} \rightarrow 10^{-4}$), individual tracks are the stubbiest.
We can account for the observed spread in color along the horizontal branch by
accepting that there is also a spread in stellar mass along this branch, bluer
stars being less massive (on the average) and less luminous than redder stars.
}  

Comparing HB models to observed globular cluster CMD's 
\citet{rood73} found that an HB that 
{\it
``\ldots is made up of stars with the same core mass and slightly varying total
mass, produces theoretical c-m diagrams very similar to those observed. \ldots
A mass loss of perhaps 0.2~M$_\odot$ with a random dispersion of several
hundredths of a solar mass is required somewhere along the giant branch.''
}
The assumption of mass loss on the red giant branch
diminished the need for very high helium
abundances.

While \citet{swgr74,swgr76} showed that HB tracks including semi-convection
covered a larger temperature range, 
\citet{swei87} noted that even with semi-convection a spread in 
mass was still necessary to explain the observations.

\citet{calo72} investigated the ZAHB locations of stars
with very low envelope masses ($\le$0.02~\Msolar) that lie along the extended or
{\bf extreme HB} (= EHB) at high effective temperatures ($>$20,000~K) and found
that they can be identified with the subdwarf B stars known in the field
\citep{gree71}. \citet{swme74} and 
\citet{ging76} studied the post-HB evolution and found that -- in contrast to
the more massive blue HB stars -- EHB models do not ascend the second
(asymptotic) giant branch (AGB), but evolve directly to the white dwarf 
domain. 

Thus our current understanding sees {\bf horizontal branch stars} as 
stars that burn helium in 
a core of about 0.5~\Msolar\ and hydrogen in a shell. The more massive the 
hydrogen envelope is the cooler is the resulting star. The masses of the
 hydrogen envelopes vary from 0.02~\Msolar\
to more than 0.2~\Msolar\ for metal-poor hot HB stars\footnote{Due to the 
higher opacities in their envelopes metal-rich HB stars are cooler than 
metal-poor ones with the same envelope mass. Therefore hot metal-rich HB 
stars must have less massive envelopes than metal-poor ones, reducing the upper 
limit to, e.g., $\approx$0.15\Msolar\ for solar-metallicity hot HB stars}.
Hot HB stars eventually evolve up the asymptotic giant branch. 
The less-massive
envelopes of the even hotter EHB stars (M$_{\rm env} \le 0.02$\Msolar,
\teff\ $>$20,000~K)
do not support hydrogen shell burning and EHB stars
do not climb the AGB, but evolve directly to the white dwarf domain and
are thus also called AGB manqu\'e stars \citep{grre90}. 
For a review on HB evolution see \citet{swei94}.
In the CMD hot horizontal branch stars populate the blue horizontal 
branch and the brighter part of the blue tail.
The transition from hot to extreme HB 
stars takes place towards the fainter part of the blue tail at $M_V 
\gtrsim$\magn{3}.

But hot horizontal branch stars are neither the brightest nor the bluest stars 
in globular clusters: Already \citet[p.\ 30]{shap30} remarked that 
{\it
``Occasionally, there are abnormally bright blue stars, as in Messier~13, but
even these are faint absolutely, compared with some of the galactic B stars''.
}
This statement refers to stars like those mentioned by \citet{barn00}
which in colour-magnitude diagrams lie above the horizontal branch and blueward
of the red giant branch (see Fig.~\ref{ag_cmd}). This is also the region where
one would expect to find central stars of planetary nebulae, which are,
however, rare in globular clusters: Until recently \citep{jamo97}
Ps1 \citep{peas28},
the planetary nebula in M~15 with its central star K~648, 
and IRAS18333-2357 in M~22 \citep{cogi89} remained the only
such objects known in globular clusters 
(see also Sect.~\ref{sec_mola98}).

Apart from analyses of individual stars like vZ~1128 in M~3
\citep[and references therein]{stst70} and Barnard~29 in M~13 
\citep{trav62,stgr68} the first
systematic work on these bright blue stars
was done by \citet{stst70b}. All stars analysed
there show close to solar helium content, contrary to the hot and extreme 
horizontal branch stars, which in general are depleted in helium 
\citep[see also Sect.~\ref{hb_metal_poor}]{hebe87,mosw00}.
Strom et al. identified the brightest and bluest UV bright
stars with models of post-AGB
stars \citep[confirming the ideas of][]{scha70} and the
remaining ones with stars evolving from the horizontal branch towards the AGB.
This means that all of the stars in their study are in the double-shell burning
stage. \citet{zine72} performed a systematic search for such stars
using the fact that they are brighter in the U band than all other cluster
stars. 
This also resulted in the name {\bf UV Bright Stars} for stars brighter
than the horizontal branch and bluer than the red giant 
branch\footnote{As
the flux maximum moves to ever shorter wavelengths for increasing
temperatures, hot UV bright stars may be rather faint not only in $V$, but
also in the $U$ band (see also Sect.~\ref{sec_mola98}).
Thus UV bright stars will appear brighter than the HB and bluer
than the red giant branch only if they are cool and/or luminous.}. 

Most of the UV bright stars found in ground based searches are cooler than
30,000~K, although theory predicts stars with temperatures up to 100,000~K
\citep[e.g.,][]{scho83,renz85}.
The ground based searches, however, are
biased towards cooler stars due to the large bolometric corrections for
hotter stars$^4$. It is therefore not surprising that space based searches in
the vacuum 
UV \citep[Ultraviolet Imaging Telescope,][]{stco97} discovered
a  considerable number of additional {\em hot} UV bright stars in a number of 
globular clusters (see also Sect.~\ref{sec_mola98}).

Space based observatories also contributed a lot of other information about hot
stars in globular clusters: Observations with the Ultraviolet Imaging 
Telescope (UIT) showed the unexpected presence of
blue HB stars in metal-rich globular clusters like NGC~362 \citep{dosh97} 
and 47~Tuc \citep{ocdo97}. At about the same
time Hubble Space Telescope (HST) observations of the core regions of globular
clusters showed long blue tails in metal-rich bulge globular clusters 
\citep{riso97}. These metal-rich globular clusters are discussed in 
more detail in Sect.~\ref{hb_metal_rich}.
The interest in hot old stars like horizontal branch and UV bright stars
has been revived and extended by the discovery of the UV excess in elliptical
galaxies \citep{cowe79,debo82} for which they
are the most likely sources 
\citep[see also Sects.~\ref{hb_metal_rich} 
and \ref{sec_mola98}]{grre90,grre99,dooc95,dorm97,brfe97}

The most recent addition to the family of hot
stars in globular clusters are the white  dwarfs found in HST observations of
M~4 \citep{rifa95,rifa97}, NGC~6752 \citep{rebr96}, NGC~6397 
\citep{pade95,copi96}, and 47~Tuc \citep{zore01},
which are discussed in Sect.~\ref{white-dwarfs}.

\section{Horizontal Branch Stars in Metal-Poor Globular Clusters
\label{hb_metal_poor}}
\subsection{Gaps and Blue Tails\label{sec-gaps}}

As mentioned in Sect.~\ref{history} the more vertical extensions of the 
blue HB (blue tails, cf. Fig.~\ref{ag_cmd}) seen in the
colour-magnitude diagrams (CMD's) of many globular clusters often
display gaps at varying
brightness. Such gaps are also known for field HB stars \citep{newe73,hehu84}.
For a list of globular clusters with blue tails see
\citet{fufe93}. \citet{cabo98} and \citet{fepa98} give comprehensive lists
of clusters that show gaps and/or bimodal horizontal 
branches\footnote{In recent deep colour-magnitude diagrams a group
of very faint blue stars ($M_V\ge$\magpt{4}{5}) shows up in some globular
clusters, e.g. NGC~2808 \citep{sodo97,walk99,bepi00},
$\omega$ Cen \citep{whro98,dcoc00}.}.
\citet{fepa98} argue that all intermediate metallicity globular clusters
([Fe/H] $\approx-1.5$) with a very long blue tail show a gap at about
18,000~K. \citet{pizo99} extend the discussion to include metal-rich globular
clusters and argue for a gap at {\em constant mass}, which --  for
differing metallicities -- will result in gaps at different temperatures.
In the following discussion we will refer to the stars along the vertical
extensions of the blue HB simply as blue tail (BT) stars and the
stars along the horizontal part of the HB (bluer than the RR Lyrae
gap) will be called blue HB (BHB) stars. Calling the stars along the 
vertical extension of the blue HB
subdwarf B stars\footnote{For analyses of field subdwarf B (sdB) stars see
\citet{hebe86,mohe90,sabe94,sake97}} \citep[e.g.,][]{baea92}
or extreme HB stars (i.e. stars with
so little hydrogen envelope that they do not burn hydrogen in a shell)
makes implicit assumptions about their physical nature and evolutionary
status that are in most cases not correct 
\citep[a point very well illustrated in Fig.~8 of][]{teco01}. 

As the gaps are not expected from canonical evolutionary scenarios various
non-canonical explanations have been suggested during the past 25 years and
some of them are given below \citep[more detailed descriptions of possible
explanations for the gaps can be found in][]{crro88,cabo98,fepa98}. 

{\it Diverging evolutionary paths}\\
The evolution away from the zero-age HB (ZAHB) could in
principle transform a uniformly populated ZAHB into a bimodal HB as stars
evolve. \citet{newe73} was the first to suggest this explanation
for the gap seen in $UBV$ photometry of field horizontal branch stars at
temperatures corresponding to $\approx$12,900~K. \citet{hehu84} suggested 
that the small gap at $\approx$20,000~K between field HBB (=horizontal 
branch B type) and sdB stars
could be explained by diverging evolution. 

Support for this idea  came from \citet{lede94}, but other calculations
show that the effect is not large enough to explain the gaps along the
horizontal branches \citep[see, e.g.,][]{dole91,cabo98}.

{\it Mass loss}\\
 \citet{dcdo96} found that bimodal horizontal branches become more probable
for increasing metallicity because the range in mass loss efficiency
required to produce an EHB star stays constant (i.e. independent of
metallicity), whereas only a very narrow range of mass loss efficiency can
produce hot HB stars at high metallicities. Thus the number of hot HB stars
is expected to decrease with increasing metallicity, opening a wide gap
between cool HB stars and EHB stars at high metallicity. 
\citet{yode00} find that mass loss {\em on} the
horizontal branch could produce extreme HB stars (like the sdB's) in very
metal-rich environments like the open cluster NGC~6791 ([Fe/H] = $+$0.5).
While these
scenarios offer good explanations for the sdB stars and the large gap
discovered in the metal-rich open cluster NGC~6791 \citep{kaud92,lisa94} 
it cannot
explain the smaller gaps seen in the mostly rather metal-poor globular
clusters. \citet{rowh97} and \citet{fepa98} also discuss variations in mass
loss on the red giant branch as possible causes for gaps along the HB.
\citet{calo99}, however, argues that HB evolution would tend to fill in
gaps in the initial ZAHB distribution if the RGB mass loss was actually
able to produce them.

{\it Differences in, e.g., [CNO/Fe], rotation etc.}\\
\citet{rocr89} suggest differences in CNO or He abundances
or rotation rates as possible causes
for the gaps. For hot HB stars a decrease in {\em CNO abundances} results
in bluer colours at a given envelope mass
(a similar effect as seen for a decrease in overall
metallicity). Increasing the {\em He abundance} in the hydrogen envelope of
a hot HB star will increase the energy production in the H-burning shell,
thereby resulting in brighter horizontal branch stars
\citep[for more details see][]{swei97b}. {\em
Rotation} would delay the helium core flash in a red giant thereby leading
to an increase in the helium core mass and more mass loss, resulting in
bluer and brighter HB stars 
\citep[see also][for a discussion of rotation and blue 
tails]{buco85,pero95,sipi00}.
Bimodal distributions in any of these
parameters may thus create gaps along the horizontal branch. 

{\it Dynamical interactions}\\
A gap would be easy to understand if the stars above and below the gap were
created by different mechanisms: If the stars below the gaps do not descend
from red giants there is no reason why they should form a smooth extension of
the sequence defined by red giants descendants. The most prominent candidates 
for such different formation mechanisms are binary interactions
like common envelope evolution, merging of stars, etc. 
\citep[for more details see][]{baea92,bail95,mohe97b}.

Such binary scenarios create stars that resemble the sdB, sdOB, and sdO
stars known from the field of the Milky Way, but not hot HB stars. The main
objection to the dynamical scenarios is that in this case the relative numbers
of red giant (RGB/AGB) to ``true'' HB stars, which gives an estimate of the
cluster's original helium abundance, would vary between clusters and
pretend varying primordial helium abundances \citep[see][]{buco85,fufe93}.
Another objection is the tight sequence in temperature and surface gravity
reported by \citet{heku86} and \citet{mohe97b}  for stars below the faint gap in
NGC~6752. \citet{crro88} cite the similar blue tails in M~15 and NGC~288,
which are dynamically very different, as argument against the production of
blue tail stars by dynamical interactions like merging. \citet{fepa97}
argue in the same way with respect to M~13 and M~3, which are dynamically
very similar but have very different HB morphologies.
\citet[NGC~2808]{bepi00} and \citet[$\omega$ Cen]{dcoc00} find no radial
gradient in the number of very faint blue stars$^5$ to blue HB stars, 
arguing against dynamical interactions as cause for the extremely faint
blue stars. \citet{teco01} on the other hand find the most pronounced blue
tail in the most metal-rich, but also densest globular cluster of their
sample, NGC~6626, which also shows indications for a higher than
usual helium content. See \citet{buco97} for a detailed discussion of the
relation between cluster density and the presence of blue tails. 

\citet{soke98} suggests that the interaction of red giants with close-in 
planets will spin-up the red giant, thereby increasing its mass loss and 
the temperature of the resulting HB star. The different fates of a planet 
inside an extended stellar envelope could then result in multimodal HB 
morphologies. So far, only 47~Tuc has been searched for planets, with 
negative results \citep{gibr00}.

{\it Atmospheric processes}\\
\citet{calo99} proposed the change from convection
to diffusion in the stellar atmospheres as an explanation for the gaps
around $(B-V)_0$ = 0. This scenario would predict chemical peculiarities in
bluer stars. \citet{grca99} suggest radiative levitation 
of heavy elements in the atmosphere as 
cause for the $u$-jump observed in many globular clusters -- a claim
which is supported by the calculations of \citet{hule00}.
A more detailed discussion of the r\^ole of
diffusion in hot HB stars can be found in Sect.~\ref{sec-diff}.

{\it Helium mixing}\\
 Helium mixing in red giants means mixing
deep enough to enrich the red giant's envelope with helium freshly produced
in the hydrogen burning shell. A red giant experiencing helium mixing will
evolve to higher luminosities, thereby losing more mass than canonically
expected and producing a hotter HB star. The helium enrichment of the
hydrogen envelope increases the efficiency of the hydrogen shell burning
and thus the luminosity of the HB star 
\citep[see][for more details]{swei97a,swei97b}. 
Different amounts of mixing in the red giant precursors
could thus produce HB stars in different temperature regimes and at the
same time explain some of the puzzling abundance distributions found in
globular cluster red giants \citep[see][for reviews, but also 
\citet{grbo01} for most recent evidence of primordial abundance 
variations]{kraf94,krsn97}. \citet{chde00} and \citet{calo01}, however,
argue that current observational results both for HB stars and red giants 
do not support the idea of helium mixing being active in globular cluster 
red giants.

{\it Statistical fluctuations}\\
\citet{cabo98} used numerous synthetic HB simulations to show quite
convincingly that at least some of the gaps may be due to statistical
fluctuations. \citet{fepa98} and \citet{pizo99}, however, report gaps at
physically similar positions (i.e. temperature or mass) in several globular
clusters, arguing against statistical fluctuations.
\subsection{Atmospheric parameters (\teff, \logg)\label{other}}
\begin{figure}[t]
\vspace*{7.6cm}
\includegraphics{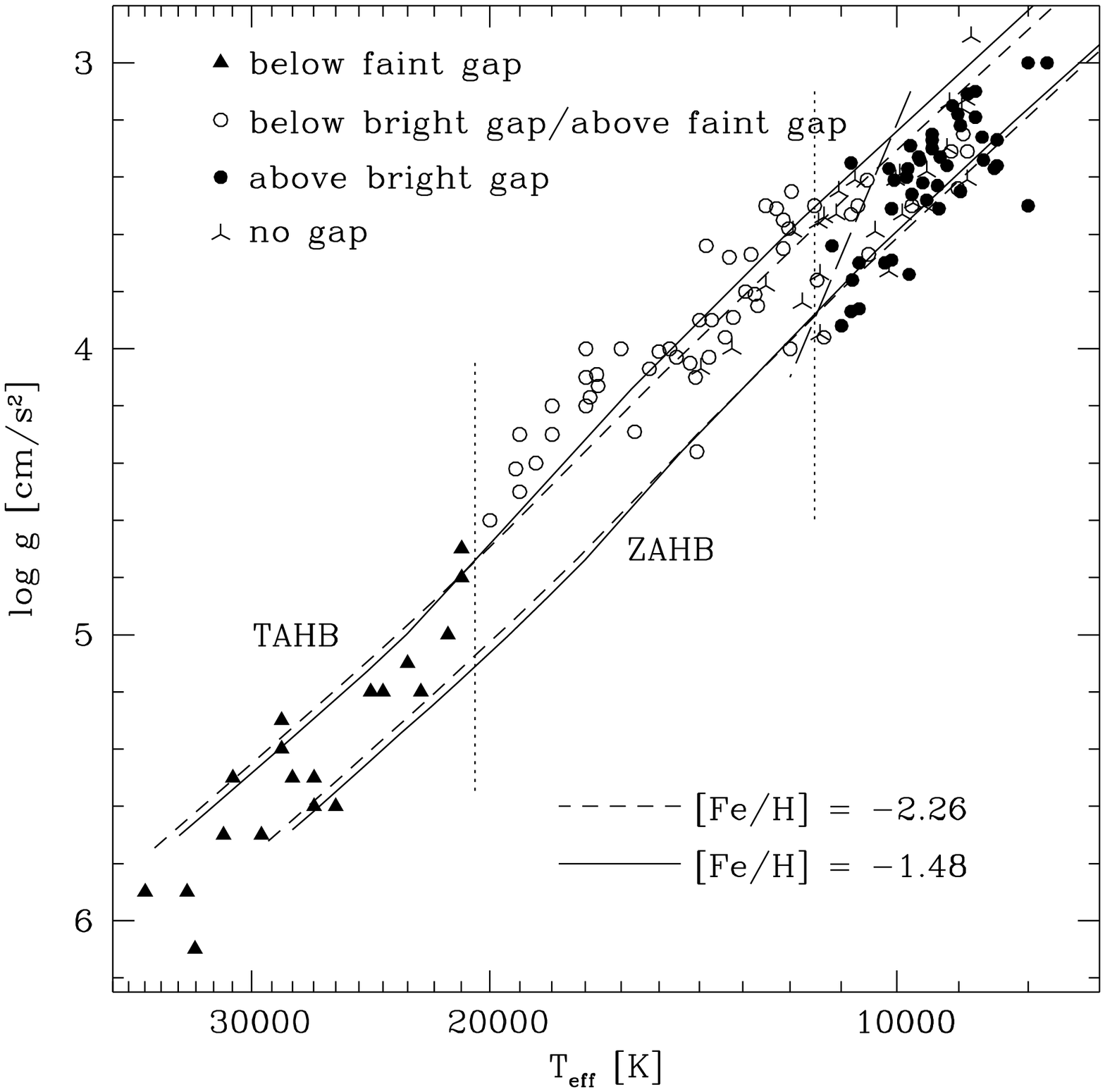}
\caption[]{The results of 
\citet[M~3, M~5, M~92, NGC~288]{crro88}, \citet[NGC~6397]{dbsc95},
\citet[M~15]{mohe95,mohe97a}, \citet[NGC~6752]{mohe97b} compared
to evolutionary tracks from \citet{doro93}. ZAHB and TAHB stand for
zero-age and terminal-age HB (see text for details). 
The long-dashed line marks the possible low-temperature gap. The 
dotted lines mark the regions of low \logg\ (see text for details). \\ 
\label{ag_plottga1}}
\end{figure}

Already early studies of hot HB stars in globular clusters showed 
discrepancies between observational results and theoretical expectations:

\citet{grdo66} mentioned that the comparison of $(c_1)_0$ vs.
$(b-y)_0$ for 50 blue HB stars in NGC~6397 to models from 
\citet{miha66} indicated low surface gravities and a mean mass of
0.3\Msolar\ (0.4\Msolar ) for solar (negligible) helium abundance, assuming
$(m-M)_0 =$ \magpt{12}{0} and \ebv\ = \magpt{0}{16}. 
{\it ``It is clear that the
accurate fixing of this parameter {\rm [log g]} is of the greatest importance
for fixing limits to the masses of the horizontal branch stars since there
seems no other way, at present, of determining them more
directly.''}
Later spectroscopic analyses of HB stars (see cited papers for details)
in globular clusters with and 
without gaps along their horizontal branches and/or blue tails
reproduced this effect (cf. Fig.~\ref{ag_plottga1}):
\citet{crro88} deal with five globular clusters, namely M~3, 
M~5, M~15, M~92, and NGC~288 (of which M 5 does not show any gap along
the BHB/BT). \citet{dbsc95} analysed BHB stars in
NGC~6397, which shows a short, horizontal blue HB. \citet{mohe95,mohe97a} 
study blue tail stars in M~15.
\citet{heku86} and \citet{mohe97b} analyse blue tail stars
in NGC~6752 which is well known for its extremely long blue tail. 

The zero-age HB (ZAHB) in Fig.~\ref{ag_plottga1} marks the position where the HB
stars have settled down and started to quietly burn helium in their cores.
The terminal-age HB (TAHB) is defined by helium exhaustion
 in the core of the HB star ($Y_C < 0.0001$). In order to allow a
better search for any common physical gaps the stars are marked by their
position relative to gaps along the HB: M~92, M~15, M~3, and NGC~288 
show a gap at 
$M_V \approx$ \magpt{0}{6} to \magpt{1}{4} (bright gap). Stars above that 
bright gap are marked by filled circles, stars below by open circles. 
M~15 and NGC~6752 
show a faint gap (or underpopulated region) at $M_V \approx$ \magn{3}. 
Stars below these faint gaps are marked by filled triangles.
NGC~6397 and M~5 show no obvious gaps (three-pointed symbols).
Fig.~\ref{ag_plottga1} shows that
the faint gap separates hot HB from EHB stars at about 20,500~K, 
which is somewhat hotter than the ``temperature'' gap for 
intermediate metallicity clusters at 18,000~K 
suggested by \citet{fepa98} and could correspond to the ``forbidden
mass'' region discussed by \citet{pizo99}. The bright gap 
roughly corresponds to the underpopulated region at 
\teff\ $\approx$10,000~K to 12,600~K (long-dashed line), although the 
distinction between stars above and below the bright gap is not as clear as 
for the faint gap. The blue tail
stars below the bright gap (and above the faint gap) are hot HB stars 
and {\em not} hot subdwarfs like the field sdB stars. 
For temperatures between 11,500~K and 20,500~K the observed
positions in the (\logg, \teff)-diagram fall mostly above the ZAHB and in
some cases even above the TAHB\footnote{\citet{croc91} finds from the analysis
of spectra for BHB stars in M~3 and M~13 that the M~3 stars cooler
than 11,200~K stay very close to the ZAHB (the one star at \teff $\approx$
12,500~K shows lower \logg ). The M~13 stars cooler than 11,200~K stay
mostly close to the ZAHB, but the majority of stars in that cluster is
hotter and shows lower \logg .}. This agrees with the finding of
\citet{sake97} that
field HBB stars show a larger scatter away from the ZAHB in \teff, \logg\
than sdB stars. 

Knowing the atmospheric parameters of the stars and the distances to the
globular clusters allows to determine masses for the stars
\citep[cf.][]{mohe94,mohe95,mohe97b,dbsc95}. While the stars in M~3, M~5,
and NGC~6752 have mean masses consistent with the canonical values, the hot
HB stars in all other clusters show masses that are significantly lower
than predicted by canonical HB evolution -- even for temperatures cooler
than 11,500~K where the stars don't deviate from the canonical tracks in
surface gravity. Scenarios like the merging of two helium-core white dwarfs
\citep{ibtu84} or the stripping of red giant cores \citep{ibtu93,tuch85}
produce low-mass stars that are either too hot (merger) or too short-lived
(stripped core) to explain the low-mass HB stars. 

Also some UV observations suggest discrepancies between theoretical expectations
and observational results: The IUE (International Ultraviolet Explorer) 
and HUT
(Hopkins Ultraviolet Telescope) spectra of M~79 \citep{alma93,dida96}
suggest lower than expected gravities and higher than expected
metallicities for hot HB stars 
\citep[but see][who do not need low surface gravities to fit 
the HUT data]{vihe99}. \citet{hich96}
find from UIT photometry of M~79 that stars bluer than $m_{152}-m_{249}$ =
\magpt{-0}{2} lie above the ZAHB, whereas cooler stars scatter around the
ZAHB. \citet[UIT data of M~13]{pabo98} find a lack of stars close to the ZAHB 
at a colour (temperature) range similar to the low \logg\ range shown in 
Fig.~\ref{ag_plottga1}. 
\citet{whro98} claim from UIT
observations that the bluest HB stars in $\omega$~Cen have lower than
expected luminosities and that a considerable number of stars lie below
the ZAHB. This is confirmed by HST observations of \citet{dcoc00} who find 
a ``blue hook'' feature at the extremely hot end of the blue tail in $\omega$ 
Cen and also several sub-ZAHB stars. These ``blue hook'' stars could be 
similar to the helium-rich sdB found in M~15 \citep{mohe97a}.
\citet{lasw96} on the other hand find good agreement between UIT
photometry of blue stars in NGC~6752 and a standard ZAHB (in position and
HB luminosity width) for $(m-M)_0$ = \magpt{13}{05} and \ebv\ =
\magpt{0}{05}. 

So far we discussed results from low to medium resolution spectra. High 
resolution spectra offer further insights into the nature of hot HB stars, 
esp. their abundances and rotational velocities, which are discussed in the 
next two sections. We'll come back to the problems described here in 
Sect.~\ref{n6752bhb_sec_par}.

\subsection{Rotational velocities\label{rotation}}

\citet{pete83,pete85a,pete85b} found from high-resolution spectroscopic
studies of blue HB stars in M~3, M~4, M~5, M~13, and NGC~288 that clusters
with bluer HB morphologies show higher rotation velocities among their HB
stars, which supports the idea that rotation affects the distribution of
stars along the HB.  However, the analysis of \citet{pero95} shows that
while the stars in M~13 (which has a long blue tail) rotate on average
faster than those in M~3 (which has only a short blue HB), the stars in
NGC~288 and M~13 show {\em slower} rotation velocities at {\em higher}
temperatures. These results are consistent with those reported for blue HB
and blue tail stars in M~13 by \citet{bedj99}, who determined rotational
velocities for stars as hot as 19,000~K \citep[considerably hotter than the 
stars analysed by][]{pero95}. They found that stars hotter than about
11,000~K have significantly lower rotational velocities than cooler stars
and that the change in mean rotational velocity may coincide with the gap
seen along the blue HB of M~13. Also the results of \citet[M~92]{comc97} and 
\citet[M15]{beco00} show that HB stars cooler than
$\approx$11,000~K to 12,000~K in general rotate faster than hotter stars.

\citet{sipi00} study theoretical models for the rotation of HB stars and find
that the observed rotation of cool BHB stars in M~13 can be explained 
if the RGB stars have rapidly rotating cores and differential
rotation in their convective envelopes and if angular momentum is redistributed 
from the rapidly rotating core to the envelope (most
likely on the horizontal branch). If, however, 
turn-off stars rotate with less than 4~km/s, a rapidly rotating core in the
main-sequence stars (violating helioseismological results for the Sun) or
an additional source of angular momentum on the RGB 
\citep[e.g. mass transfer in
close binaries or due to planets as described by][]{soha00} are required to
explain the rotation of BHB stars. The change in rotation rates towards
higher temperatures is not predicted by the models but could be understood
as a result of gravitational settling, which creates a mean molecular
weight gradient, that then inhibits angular momentum transport in the star.
\citet{swei01} suggests that the weak stellar wind invoked to 
reconcile observed abundances in hot and extreme HB stars with diffusion 
calculations (cf. Sect.~\ref{sec-diff}) 
could also carry away angular momentum from the
surface layers and thus reduce the rotational velocities of these stars.

\citet{soha00} argue that the distribution of rotational velocities along
the HB can be explained by spin-up of the progenitors due to interaction
with low-mass companions, predominantly gas-giant planets, in some cases
also brown dwarfs or low-mass main-sequence stars (esp. for the very hot
extreme HB stars). The slower rotation of the hotter stars in their
scenario is explained by mass loss {\em on} the HB, which is accompanied by
efficient angular momentum loss. 
This scenario, however, does not explain the sudden change in 
rotational velocities and the coincidence of this change 
with the onset of radiative 
levitation. 

\subsection{Atmospheric abundances\label{sec-diff}}
It has been realized early on that the blue HB and blue tail stars in
globular clusters show weaker helium lines than field main sequence B stars
of similar temperatures: \citet[NGC~6397]{sero66}; 
\citet[M~5, M~13, M~92]{grmu66}; 
\citet[M~13, M~15, M~92]{sarg67}. \citet{grtr67} already suggested
diffusion to explain this He deficiency.

\citet{miva83} performed the first theoretical study of
diffusion effects in hot and extreme
horizontal branch stars. Using the evolutionary
tracks of \citet{swgr76} they found for the metal-poor 
models that {\it ``in most of each
envelope, the radiative acceleration on all elements {\rm (i.e. C, N, O,
Ca, Fe)} is much larger than gravity which is not the case in main-sequence
stars.''} The elements are thus pushed towards the surface of the 
star. Turbulence affects the different elements to varying extent, but generally
reduces the overabundances\footnote{\citet{mich82} and 
\citet{chmi88} showed that meridional circulation can prevent
gravitational settling and that the limiting rotational velocity decreases
with decreasing \logg. 
\citet{beco00} note that two of the HB stars 
hotter than 10,000~K show higher rotational velocities and much smaller 
abundance deviations.}. 
Models without turbulence and/or mass loss (which may reduce the effects of
diffusion) predict stronger He depletions than are
observed. A weak stellar wind could alleviate this discrepancy 
\citep[discuss this effect, albeit for hotter 
stars]{hebe86,mibe89,foch97,unbu98}.

The extent of the predicted abundance variations varies with effective
temperature, from none for HB stars cooler than about $5800 \pm 500$K (due
to the very long diffusion timescales) to 2 -- 4 dex in the hotter stars
(the hottest model has \teff\ = 20,700~K) and also depends on the element 
considered.
The overabundances in the two hottest models (12,500~K and 20,700~K) are
limited to 3 dex for relatively abundant elements by the saturation of
lines. Less abundant elements like P, Eu, Ga could show much larger
overabundances before their lines saturate (up to 5 dex for original values
of [M/H] = $-$2). 

Observations of BHB and BT stars in globular clusters support the
idea of diffusion being active above a certain temperature:

Abundance analyses of blue HB stars cooler than 11,000~K to 12,000~K in 
general show no deviations from the globular cluster abundances derived 
from red giants: \citet[NGC~6397]{glde86}, \citet[NGC~6752]{glmi89},
\citet[M~4, NGC~6397]{lamc92}, \citet[M~92]{comc97}
\citet[M~13]{beco99}, \citet[M~15]{beco00}, \citet[NGC~6752]{pero00}.
For stars hotter than 11,000~K to 12,000~K, however, departures from the 
general globular cluster abundances are found, e.g.\ iron enrichment to 
solar or even super-solar values and strong helium 
depletion: \citet[NGC~6752]{glmi89}, \citet[M~13]{beco99}, 
 \citet[NGC~288, M~13]{pero95},
\citet[NGC~6752]{mosw00}, \citet[M~15]{beco00}, 
\citet[NGC~6752]{pero00}.
This agrees with the finding of \citet{alma93} and \citet{vihe99}
that solar metallicity model atmospheres are required 
to fit the UV spectra of M~79. 

All this evidence supports the recent suggestion of \citet{grca99} that the
onset of diffusion in stellar atmospheres may play a r\^ole in explaining
the jump along the HB towards brighter $u$ magnitudes at effective
temperatures of about 11,500~K. This jump in $u, u-y$ is seen in all CMD's
of globular clusters that have Str\"omgren photometry of sufficient
quality\footnote{\citet{bepi00} report a $U$ jump for NGC~2808 and
\citet{masp01} detect it in their $UBV$ photometry of M~5.}. The observed HB
stars return to the theoretical ZAHB at temperatures between 15,000~K and
20,000~K \citep[Fig.~1]{grca99}. The effective temperature of the jump is
roughly the same for all clusters, irrespective of metallicity, central
density, concentration or mixing evidence, and coincides with the apparent
gap in \teff, \logg\ seen in Fig.~\ref{ag_plottga1} at \teff\
$\approx$10,000~K to 12,000~K. This coincides with the region where surface 
convection zones due to hydrogen and \ion{He}{1} ionization disappear in HB 
stars \citep{swei01}.

Radiative levitation of heavy elements decreases the far-UV flux and by
backwarming increases the flux in $u$. \citet{grca99} show that the use of
metal-rich atmospheres ([Fe/H] = $+0.5$ for scaled-solar ATLAS9 Kurucz
model atmospheres with $\log \epsilon_{Fe,\odot} = 7.60$) improves the
agreement between observed data and theoretical ZAHB in the $u, u-y$-CMD at
effective temperatures between 11,500~K and 20,000~K, 
but it worsens the agreement between theory and
observation for hotter stars in the Str\"omgren CMD of NGC~6752 (see their
Fig.~8). Thus diffusion may either not be as important in the hotter stars
or the effects may be diminished by a weak stellar wind.
 
The gap at $(B-V)_0 \approx 0$ discussed by 
\citet[see Sect.~\ref{sec-gaps}]{calo99} 
is not directly related to the $u$-jump 
as it corresponds to an effective
temperature of about 9000~K and is also not seen in every cluster
(which would be expected if it were due to an atmospheric phenomenon). 
The gap at \teff\ $\approx$ 13,000~K seen in the $c_1, b-y$ diagram of
field horizontal branch stars \citep{newe73,negr76} may be related to the 
$u$-jump as the $c_1$ index contains $u$. 

The abundance distribution within a stellar atmosphere 
influences the temperature stratification and thereby 
the line profiles and the flux distribution of the emergent 
spectrum. A deviation in 
atmospheric abundances of HB stars from the cluster metallicity due to
diffusion would thus affect their line profiles and flux distribution. Model 
atmospheres calculated for the cluster metallicity may then yield wrong 
results for effective temperatures and surface gravities when compared to 
observed spectra of HB stars. Self-consistent model atmospheres taking into 
account the effects of gravitational settling and radiative levitation are, 
however, quite costly in CPU time and have started to appear only quite 
recently for hot stars \citep{drwo99,hule00}. 

\subsection{Atmospheric parameter revisited\label{n6752bhb_sec_par}}

Analysis of a larger sample of hot and extreme HB stars in NGC~6752
\citep{mosw00} showed that the use of model atmospheres with solar or 
super-solar abundances removes much of the deviation from canonical tracks 
both in \teff, \logg\ and \teff, mass for hot HB stars discussed in 
Sect.~\ref{other}. However, some 
discrepancies remain, indicating that the low \logg, low mass problem cannot be
completely solved by scaled-solar metal-rich atmospheres 
\citep[which {\em do} reproduce the $u$-jump reported by][]{grca99}. 
As \citet{miva83} noted diffusion will not necessarily enhance all heavy
elements by the same amount and the effects of diffusion vary with
effective temperature. Elements that were originally very rare may be
enhanced even stronger than iron \citep[see also][where P and Cr
are enhanced to supersolar abundances]{beco99}.
The question of whether diffusion is the
(one and only) solution to the ``low gravity'' problem cannot be
answered without detailed abundance analyses to determine the actual
abundances and model atmospheres that allow to use non-scaled solar
abundances \citep[like ATLAS12][]{kuru92}. 

\subsection{Where do we stand?\label{hb_final}}

The spectroscopic analyses of BHB and blue tail stars in globular clusters
suggest that the faint gap or underpopulated region at $M_V
\approx$\magn{3} can be identified with the transition from hot to extreme
HB stars, while the bright gap is probably caused by the onset of radiative
levitation in the atmospheres of the hot HB stars. While the sudden change in
rotational velocity at the bright gap is not yet understood the good
agreement of spectroscopic results (accounting for diffusion) with
canonical evolution makes several non-canonical scenarios discussed in
Sect.~\ref{sec-gaps} appear unlikely: Helium mixing, rotation 
and high primordial helium abundance would all increase the luminosities of 
the hot HB stars \citep[resulting in lower \logg, but canonical
masses, see][]{crro88,swei97a}. Currently, however, stars with low 
\logg\ show also low masses \citep{mosw00}, suggesting deficiencies in the 
analysis rather than non-canonical evolutionary effects as cause.
Dynamical interactions are unlikely to produce the tight sequence
of stars in the \teff, \logg-diagram. These statements, however,  are
currently valid only for those (intermediate metallicity and metal-poor)
globular clusters where spectroscopic analyses of
blue tail/blue HB stars exist. More spectroscopic analyses, esp. in more 
metal-rich clusters, would help to verify the suggestion of \citet{pizo99} 
that the faint gap corresponds to a ``forbidden'' mass (which would result 
in cooler gap temperatures in more metal-rich globular clusters).

Still unexplained, however, are the low masses found for cool blue HB stars
(which are not affected by diffusion) in, e.g., NGC~6397 and M~92. For those
stars a longer distance scale to globular clusters would reduce the
discrepancies. Such a longer distance scale has been suggested by several
authors using {\sc Hipparcos} results for metal-poor field subdwarfs to
determine the distances to globular clusters by fitting their main sequence
with the local subdwarfs \citep[see][for an overview of the {\sc Hipparcos}
results]{reid99}. \citet{cagr00} present an extensive and excellent
discussion of various globular cluster distance determinations and the zoo
of biases that affect them. It is interesting to note that for M~92 and
NGC~6397 the new distance moduli are \magpt{0}{3} -- \magpt{0}{6} larger
than the old ones, thereby greatly reducing the mass discrepancies
\citep[see also][]{hemo97}. The results of spectroscopic analyses of BHB
stars (cooler than 11,000~K to 12,000~K) in globular clusters therefore
favour the longer distance scale \citep{moeh99}\footnote{\citet{detu97},
however,
report that {\sc Hipparcos} parallaxes for field HBA stars still yield
masses significantly below the canonical mass expected for these objects.}.

\section{Horizontal Branch Stars in Metal-Rich Globular
Clusters\label{hb_metal_rich}}

So far we have dealt with blue HB and blue tail stars in metal-poor ([Fe/H]
$<-1$) globular clusters. As mentioned in Sect.~\ref{history}  the HB
morphology correlates with metallicity, i.e. HB stars in metal-rich
globular clusters will populate mainly the cool regions of the HB because
for a given mass of the hydrogen envelope the resulting effective
temperature decreases with increasing metallicity. The detection of sdB/sdO
candidates in the metal-rich open clusters NGC~188 ([Fe/H] $\approx 0$) and
NGC~6791 ([Fe/H] $\approx+0.5$) by UIT \citep{labo98} and
optical photometry \citep{kaud92}, followed by
the spectroscopic verification of sdB stars in NGC~6791 
\citep{lisa94} proves, however, that at least extreme HB stars can be produced
also in metal-rich systems \citep[see also][for theoretical scenarios]{dcdo96}.

\begin{figure}[!ht]
\vspace*{7.2cm}
\includegraphics{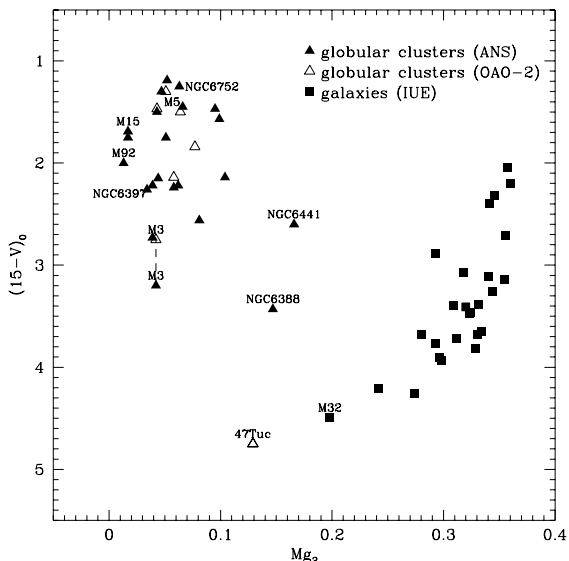}
\caption[$(15-V)_0$ vs. Mg$_2$ plot]
{UV-visual colour $(15-V)_0$ vs. metallicity index Mg$_2$ for globular
clusters and elliptical galaxies 
\citep[adapted from][$15$ being the brightness at 1500~\AA]{dooc95}. 
The metal-poor globular clusters discussed in Sect.~\ref{hb_metal_poor}
and the ``transition'' objects between globular clusters and elliptical 
galaxies are identified. \label{uv_gc_ellip}} \end{figure} 

UV observations of elliptical galaxies, which are in general even more
metal-rich than metal-rich globular clusters (based on the strength of the
Mg$_2$ index, cf. Fig.~\ref{uv_gc_ellip}) showed that such old, metal-rich 
systems contain hot stars
\citep{bube88}. Stellar evolution models yield the maximum
lifetime UV output for EHB stars with envelope masses M$_{\rm env} \le
0.02$\Msolar \citep[see also][]{grre90,grre99}, while post-AGB
stars do not live long enough at high temperatures to play a significant
r\^ole for the UV flux. Further evidence in support of hot subdwarfs
as cause for the UV excess in elliptical galaxies is provided by 
\citet{brfe97}: Their analysis of HUT
spectra of 6 elliptical and S0 galaxies shows that models with
super-solar metal and helium abundances provide the best fit to the flux
distribution of the observed spectra and that EHB stars are required in all
fits. Most absorption line features (of C, N, Si, i.e. light elements),
however, are consistent with [M/H] = $-1$, in contrast to the energy
distribution. This may be due to diffusion in the atmospheres of the EHB
stars (see Sect.~\ref{sec-diff}). 

\citet{dooc95} present a thorough discussion of the
observational evidence for UV excess in elliptical galaxies and compare the
galaxy data to those obtained for globular clusters. Comparing the
UV-visual colour $(15-V)_0$ for galaxies and globular clusters to the
Mg$_2$ metallicity index (see Fig.~\ref{uv_gc_ellip}, $15$ being the
observed brightness at 1500~\AA) they find that while the globular clusters
and the galaxies occupy distinct ranges in Mg$_2$ they overlap in
$(15-V)_0$ with the globular clusters being bluer on average. The region
between globular clusters and galaxies in Mg$_2$ is occupied by the
metal-rich globular clusters 47~Tuc, NGC~6388,
NGC~6441,
and the small elliptical galaxy M~32 \citep[see][for far-UV HST observations
of this galaxy]{brbo00}. The discovery of hot stars in the
metal-rich ``transition'' globular clusters (see Sect.~\ref{history})
is thus of special
interest as analyses of these stars may provide additional information on
the nature of the UV excess in elliptical galaxies.

\citet{rimi93} analysed IUE spectra of the cores of 11 disk globular
clusters. The surface light distribution in these spectra becomes more
concentrated towards shorter wavelengths for the clusters with the highest
UV fluxes. The UV colours of the metal-rich globular clusters NGC~6388,
NGC~6441, NGC~6624, NGC~6637 are almost as blue as those of of metal-poor
globular clusters (see Fig.~\ref{uv_gc_ellip}).
The IUE observations of NGC~6637 and NGC~6624 could be
explained by one post-EHB star or a few EHB stars,
while for NGC~6441, which shows a rise in UV flux towards shorter
wavelengths (similar to elliptical galaxies), post-HB stars are the most
likely sources. The ratio $L_{UV}/L_{total}$ of the clusters showing high
far-UV fluxes agree very well with those seen in elliptical galaxies,
whereas that of NGC~6388, which shows a flat UV spectrum (best explained by
blue HB stars), is one order of magnitude lower. 47~Tuc does not show any
evidence for stars hotter than blue stragglers within the IUE aperture. 

Some years later \citet{riso97} discovered the first well
populated blue tails in metal-rich globular clusters from WFPC2 photometry
of the cores of NGC~6388 and NGC~6441. Most surprisingly, the HB stars at
the top of the blue tail are roughly \magpt{0}{5} brighter in $V$ than
the red HB ``clump," which is strongly sloped as well. The slight 
HB tilt ($\Delta V \approx$\magpt{0}{1}) expected for metal-rich globular 
clusters due to the variation in bolometric correction for metal-rich BHB 
stars \citep{brca99} is much smaller than the observed slope.
Differential reddening alone is probably not the cause of this additional 
slope \citep{piot97,swca98,lari99}. 
WFPC2 photometry of the core of 47~Tuc obtained within the
same programme does not show any evidence for a blue HB or a blue tail nor
any slope along its red HB. \citet{lari99} verify the slope
of the blue HB and red clump in NGC~6441 and
recent analyses of RR Lyrae variables in NGC~6388 and 
NGC~6441 \citep{lari99,prsm99} 
strongly indicate that the RR Lyrae stars of these
globular clusters are substantially brighter than canonical models would
predict. 

\citet{ocdo97} detected about 20 hot stars on the UIT far-UV image of 47~Tuc,
which they identify with those producing the UV upturn in elliptical
galaxies. Their number, however, is too small to produce a significant UV
upturn in 47~Tuc. The much larger field of the UIT accounts for the
different results of UIT vs. IUE and WFPC2 observations. The small number of
hot stars in 47~Tuc agrees with the result of \citet{rode99} that only
about 7\% of the mid-UV light of 47~Tuc comes from stars hotter than about
7,500~K (most of which are probably blue stragglers). 

\citet{dosh97} find evidence for hot stars in
NGC~362 from UIT observations. While this globular cluster is not
metal-rich, its HB morphology is too red for its metallicity. Together with
NGC~288, which has a predominantly blue HB at a similar metallicity, it
forms a second-parameter pair of globular clusters (meaning that an
additional parameter besides metallicity is necessary to explain the
difference in HB morphology between these two clusters). 

What are the possible origins for the hot stars in these four globular
clusters?

{\it High mass loss tail:}\\
\citet{dosh97} and \citet{ocdo97} suggest that the hot stars in NGC~362 and
47~Tuc are simply the high mass-loss tail of the red HB distribution. 
A high mass loss tail can most probably not explain the
much more numerous blue stars in NGC~6388 and NGC~6441.
Moreover, increasing RGB mass
loss moves an HB star blueward in the $V$,~$B-V$ plane but does not
increase its luminosity (the same holds true for an increase in age). 

{\it Dynamical interactions:}\\
\citet{bail95} has reviewed the binary evolution scenarios which
could yield hot subdwarf stars in globular clusters (see also 
Sect.~\ref{sec-gaps}). Binary evolution could
be a valid explanation for 47~Tuc and NGC~362 although it is puzzling that
the center of 47~Tuc (where interactions should be most pronounced) does
not show any evidence for hot stars, whereas the core of NGC~362 shows a 
concentration of hot stars \citep{dosh97}. It is, however, 
not yet clear whether 
the hot stars in the core of NGC 362 are HB stars or extreme blue stragglers.

If dynamical interactions created the hot HB stars in NGC~6388 and NGC~6441
these stars should be more centrally concentrated than the RGB stars which
is not evident in the HST data. One should note, however, that
\citet{lari99} find a much less pronounced blue tail in the outer regions
of NGC~6441 (where of course the contamination by the field bulge
population is much stronger) and suggest that the blue HB/blue tail stars
are more centrally concentrated than the red clump stars. However, binaries 
cannot explain the slope of the HB seen in NGC~6388 and NGC~6441. 

{\it Spread in metallicity:}\\
This scenario was first discussed by \citet{piot97} to explain the sloped
HB's found in NGC~6388 and NGC~6441. Model calculations by \citet{swei01}
show that the metal-poor end ([Fe/H] = $-$2.3) of the zero-age HB (ZAHB) 
for these variable metallicity tracks is
about \magpt{0}{4} more luminous at the top of the blue tail
than the canonical ZAHB for [Fe/H] = $-$0.5. In this case NGC~6388 and NGC~6441
might be metal-rich analogues of $\omega$ Cen, the only other GC known to
show a spread in metallicity. 

Two of the mechanisms discussed in Sect.~\ref{sec-gaps} 
may also produce hot HB stars and a 
sloped HB in metal-rich globular clusters: Both {\em rotation} and {\em 
helium mixing} can create brighter and hotter HB stars. 
Rotation and/or mixing strong enough to produce the observed slope of the 
HB in NGC~6388 and NGC~6441 would at the same time produce a considerable 
number of hot stars.

While a high primordial helium abundance can also explain a sloped HB
together with a blue tail in a metal-rich globular cluster
\citep{cadf96,swca98}, this scenario also predicts a much larger value for
the number ratio $R$ (= HB/RGB) than the value recently obtained
by \citet{lari99} for NGC~6441. 

\citet{mola00}
analysed hot HB star candidates in 47~Tuc and NGC~362: 
Three of the four blue HB stars analysed
in 47~Tuc and three of the eight observed in NGC~362 are probably members
of the clusters and their parameters and masses (except for one
spectroscopic/photometric binary in 47~Tuc, which cannot be properly
analysed) agree very well with canonical evolutionary tracks. 

The three spectroscopically verified hot HB stars in 47 Tuc are much hotter
(10,000 K $<$ \teff $<$ 15,000 K) than the rest of the HB population, which
is (except for the single RR Lyr V9) entirely redward of the instability
strip\footnote{A fourth probable hot star member of 47 Tuc is UIT-14, which
is only $1.7'$ from the cluster center, and  for which the IUE spectrum
obtained by \citet{ocdo97} indicates \teff $\approx$ 50,000~K.}. The small
number of hot HB stars\footnote{\citet{kakr97} find only 2 candidates for
blue tail stars in 47~Tuc \citep[the fainter of which is very similar to
the SMC star MJ8279 discussed by][]{mola00}} in 47~Tuc, and their high
temperatures, point to a scenario in which they have a different physical
origin than the dominant red HB population (e.g. 
binary interactions, although the lack of central concentration remains a
strong caveat for this scenario). 

As the separation between the hot and cool HB stars in NGC~362 is much
smaller, it is plausible that the blue HB stars arise from a small
percentage of red giants with unusually high mass loss. The three probable
member stars in NGC~362 are all located within 2\bmin5 of the cluster
center, while the remaining five stars \citep[probably members of the SMC,
for more details see][]{mola00} are all more than 3\bmin5 from the center.
It would be interesting to study the stellar parameters of the hot stars in
the core region, where also the relative SMC contamination should be much
lower. 

The atmospheric parameters derived for the hot HB stars in NGC~6388 and
NGC~6441 \citep{mosw99} on the other hand
 place the studied stars preferentially 
{\em below} the canonical
ZAHB. The derived gravities for most stars are {\em significantly} larger
than those predicted by the non-canonical tracks (rotation, helium-mixing)
that reproduce the upward sloping horizontal branches. 

A spread in metallicity, which requires the blue tail stars to be 
metal-poor, would reduce the discrepancies found by \citet{mosw99}: The 
authors relied on the equivalent width of the \ion{Ca}{2} K line to place the 
analysed stars on the hot side of the Balmer maximum. A reduction in 
metallicity would reduce the expected equivalent width of the \ion{Ca}{2} K 
line 
also for temperatures below 9,000~K to values consistent with the observed 
ones (including the high reddening of these clusters). 
If the cool solutions were chosen all stars except one 
end up close to the ZAHB computed for varying metallicity and the problem 
of the high gravities vanishes.

In summary one can state that hot HB stars in metal-rich globular clusters
with few such stars (47~Tuc, NGC~362) show parameters in agreement with
canonical evolution (i.e. high mass loss tail), although binary evolution
may play a r\^ole. The numerous hot HB stars in NGC~6388 and NGC~6441,
however, can currently be best explained by a spread in metallicity, 
accompanied by canonical evolution.

\section{UV Bright Stars in Globular Clusters\label{sec_mola98}}
As mentioned in Sect.~\ref{history} UV bright stars 
have originally been
defined as stars brighter than the horizontal branch and bluer than red
giants \citep[see also Fig.~\ref{ag_cmd}]{zine72},
that are brighter in $U$ than any other cluster
star. 

\citet{zinn74} observed spectra of 38 optically selected UV bright stars in 8
globular clusters. He found that -- at a given age and metallicity --
different HB morphologies result in different UV bright
star populations: The presence/absence of ``supra-HB'' stars is 
correlated with the presence/absence of hot HB stars in M~13,
M~15, and M~3. This agrees with the theoretical expectation that hot HB
stars evolving away from the HB show up as ``supra-HB'' stars. The more
luminous UV bright stars 
in all three globular clusters are consistent with post-AGB
tracks. Also the existence of a planetary nebula and the presence of red HB
stars in M~15 (which is unusual for such a metal-poor globular cluster) are
linked to each other: The red HB stars in M~15 have masses of 0.8 --
0.9\Msolar, which favour the creation of planetary nebulae 
(compared to less massive BHB or blue tail stars). \citet{scho83} discusses the
theoretical evolution of post-AGB stars with special emphasis on the
production of planetary nebulae: The 0.546\Msolar\ model, which leaves the
AGB before thermal pulses start (post-early AGB), evolves so slowly that
its age at 30,000~K (the temperature for planetary 
nebula ionization) exceeds the age of the oldest known planetary nebulae.
Thus the lower mass limit for central stars of planetary nebulae is
taken to be 0.55\Msolar . 

The search for UV bright stars in globular clusters continued and
\citet{hane83} list 29 globular clusters with 23 (11) UV bright stars 
bluer than
$(B-V)_0 = 0$  that are definite (probable) cluster members. \citet{debo85}
used IUE spectra of 10 hot UV bright stars 
in 7 globular clusters to estimate their
contribution to the integrated UV light of the respective globular
clusters: hot post-AGB stars contribute less than 3\% to the total cluster
light at 3300\AA, increasing to about 15\% at 1500\AA\ and further
increasing towards even shorter wavelengths. \citet{debo87} gives a
compilation of 45 luminous hot UV bright stars ($M_V < 0$, $(B-V)_0 <0.2$)
in 36 globular clusters.

Hot post-(extreme)HB and post-(early) AGB stars do not necessarily fulfil the
original definition of UV bright stars: As stars get hotter the maximum of
their flux distribution moves to ever shorter wavelengths and especially
the less luminous UV bright stars evolving away from the extreme HB can be
quite faint at visual and near-UV wavelengths. The early lists of hot
UV bright stars 
are thus certainly incomplete as they are based on optical searches,
which favour luminous hot UV bright stars and are also limited in their
spatial coverage due to crowding in the cluster cores. As hot UV bright
stars shine up in far-UV images of globular clusters the Ultraviolet
Imaging Telescope \citep[UIT,][]{stco97} was used to obtain ultraviolet
($\sim1620$~\AA) images of 14 globular clusters. The solar-blind detectors
on UIT suppress the cool star population, which allows UV-bright stars to
be detected into the cluster cores, and the $40'$ field of view of UIT is
large enough to image the entire population of most of the observed
clusters. Thus the UIT images provide a complete census of the hot
UV-bright stars in the observed clusters, which is well suited to test
post-(extreme)HB and post-(early) AGB evolutionary tracks. Such a test is
especially important as hot UV bright stars probably make a significant
contribution to the UV-upturn observed in elliptical galaxies
\citep{grre90,dooc95,dorm97,brfe97,grre99,brbo00}. 

The need for further information on these
evolutionary stages is also illustrated by the results of \citet{jamo97}
for planetary nebulae in globular clusters. In their \ion{O}{3}
imaging survey of 133 globular clusters they found only four planetary
nebulae, two of which were previously known (Ps1 in M~15 and IRAS
18333-2357 in M~22, cf. Sect.~\ref{history}). 
Based on the planetary nebula luminosity function for
metal-poor populations they expected to find 16 planetary nebulae in their
sample. However, their \ion{O}{3} search may have missed some old, faint
planetary nebulae. And -- even more important -- their assumption that all
stars in a globular cluster will eventually go through the AGB phase is not
valid for globular clusters like NGC 6752, where about 30\% of the HB
population consist of EHB stars (with \teff\ $>$ 20,000~K), which evolve
into white dwarfs without ever 
passing through the thermally pulsing AGB phase. While such globular clusters
are expected to be deficient in post-AGB stars, they
should show a substantial population of less luminous 
(1.8 $<$ \logLL\ $<$ 3) 
UV-bright stars, which can be either post-EHB stars or post-early AGB
stars, neither of which would produce a planetary nebula. 

\begin{figure}[ht]
\vspace{7.6cm}
\includegraphics{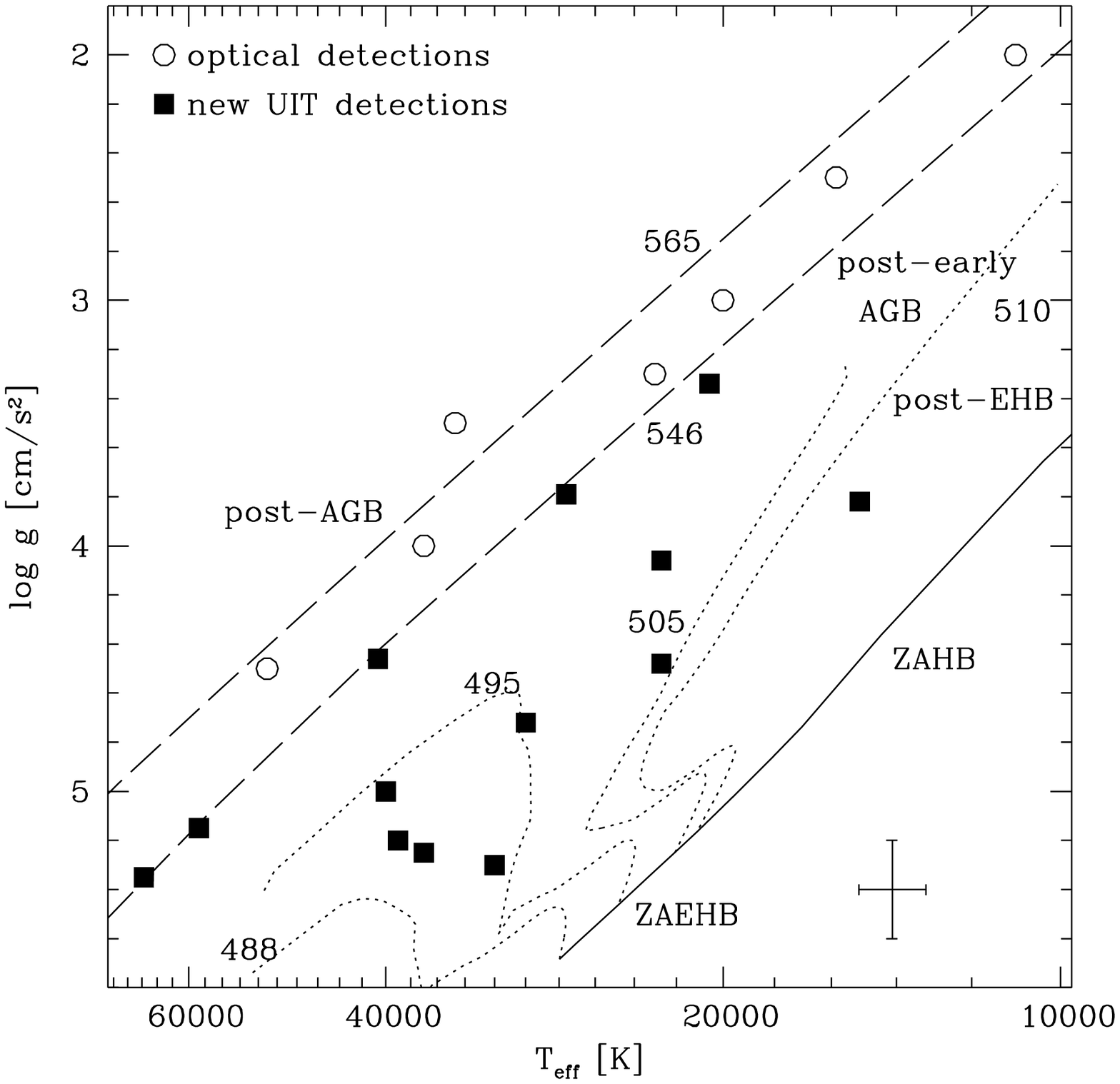}
\caption[Atmospheric parameters of the UV bright stars compared to
evolutionary tracks]
{The atmospheric parameters of hot UV bright stars compared to evolutionary
tracks. The solid and dotted lines mark the ZAHB and post-ZAHB
evolutionary tracks for [Fe/H] = $-$1.48 \citep{doro93}. The dashed lines
give post-AGB (0.565~\Msolar) and post-early AGB (0.546~\Msolar) tracks
\citep{scho83}. All tracks are labeled with the mass of the stars in units
of 10$^{-3}$\Msolar. The filled symbols mark UV bright stars identified as 
such only
by UIT, while the open symbols mark UV bright stars already known from optical
searches (see text for references). \label{eso96_tg}} 
\end{figure}

All this emphasizes the need for spectroscopic analyses of hot UV bright
stars to compare their parameters to evolutionary calculations. Most
analyses so far, however, have been limited to the use of IUE spectra.
While IUE spectra allow a good determination of \teff\ for hot stars they
are not very suitable to determine \logg\ \citep[see][]{cafu95}. 
Analyses that also used hydrogen lines (line
profile fits or equivalent widths) or the shape of the far-UV 
continuum were performed for eight optically selected 
hot UV bright stars (in some cases only the most recent analysis is given):
M22~II-81 \citep{glde85}; NGC6712-C49 \citep[only lower limit for
\teff]{reca80}; NGC~6397~ROB162 \citep{heku86b}; NGC~1851~UV5, M~3~vZ1128
\citep{dida94}; 47~Tuc~BS \citep{dida95}; M13 Barnard~29 \citep{codu94},
$\omega$~Cen~ROA5139 \citep{mohe98b}.
\citet[ground-based observations, ten stars]{mola98} and \citet[HST
observations, three stars]{wayne} observed and analysed 
spectra of UV-bright stars identified as such
solely on the UIT images.
The derived effective temperatures and gravities of all these stars are
plotted in Fig.~\ref{eso96_tg}, along with evolutionary tracks.

Obviously the dominance of post-AGB stars among optically selected hot UV
bright stars is due to heavy bias of the selection towards the most
luminous stars. The analysis of optically selected hot UV bright stars thus
gives a wrong impression of the importance of the various evolutionary
phases that contribute to the UV flux of old stellar populations. 
The lack of classic post-AGB stars among hot UV bright stars in globular 
clusters may be understood
from the different lifetimes: The lifetime of Sch\"onberner's post-early
AGB track is about 10 times longer than his lowest mass post-AGB track.
Thus, even if only a small fraction of stars follow post-early AGB tracks,
those stars may be more numerous than true post-AGB stars. Due to their
relatively long lifetime, post-early AGB stars are also unlikely to be observed
as central stars of planetary nebulae (see above).

Theoretical
simulations would be useful to determine whether the relative populations of
post-AGB and post-early AGB stars can be accommodated using existing post-HB
evolutionary tracks or if additional process (e.g. additional mass loss) 
are necessary. Possible discrepancies are indicated by \citet{lasw96}, who 
find only 4 post-EHB stars in UIT observations of NGC~6752, whereas 11 
would be expected.

\section{White dwarfs in globular clusters\label{white-dwarfs}}

White dwarfs are the final stage of all low-mass stars (like those
discussed so far) and globular clusters should thus contain lots of them. 
However,
these stars managed to evade detection until recently photometric white dwarf
sequences were discovered in four globular clusters by
observations with HST \citep{pade95,rifa95,rifa97,copi96,rebr96,zore01}.
These sequences not only allow to verify time scales for the evolution of
low-mass stars, but also offer an independent way to determine distances to
globular clusters, as suggested by \citet{rebr96}: The basic idea is to fit
the white dwarf cooling sequence of a globular cluster to an appropriate
empirical cooling sequence of local white dwarfs with well determined
trigonometric parallaxes. The procedure is analogous to the classical main
sequence fitting but has two main advantages: White dwarfs have -- due to
diffusion -- very simple atmospheres that are either hydrogen-rich (DA) or
helium-rich (DB/DO), independent of their original metallicity. Thus one
can avoid the problem to find local calibrators with the same metallicities
as the globular cluster stars. In addition, white dwarfs are locally much
more abundant than metal-poor subdwarfs, thus enlarging the reference
sample. 

Photometric observations alone, however, are not sufficient to select the
appropriate local calibrators: Hydrogen-rich DA's and helium-rich DB's can
in principle be distinguished by their photometric properties alone in the
temperature range $10,000\,K \leq T_{\mathrm{eff}} \leq 15,000$\,K
\citep{bewe95}. \citet{rebr96} classified two white dwarfs in NGC\,6752 as
DB's by this method and \citet{rifa97} speculate that the brightest white
dwarf in M\,4 (V=22.08) might be a hot (27,000K) DB star. However, without
a spectral classification, those stars could also be high-mass DA white
dwarfs, possibly a product of merging. 

Also, the location of the white dwarf cooling sequence is highly sensitive
to the white dwarf mass. \citet{rebr96} argued that the white dwarf masses
in globular clusters are constrained to the narrow range 0.51\Msolar\
$\leq$ M$_{\rm WD} \leq$ 0.55\Msolar, but some systematic differences
between clusters are obvious: At a given metallicity some globular clusters
(e.g.\ NGC\,6752) possess very blue horizontal branches whose low-mass
extreme HB stars evolve directly to low mass C/O white dwarfs (bypassing
the AGB) and shift the mean white dwarf mass closer to 0.51\Msolar. Other
clusters show only red HB stars, which will evolve to the AGB and form
preferably white dwarfs with masses of $\approx$0.55\Msolar. In addition,
low mass white dwarfs (M$<$0.45\Msolar) with a degenerate He core (instead
of the ``normal'' C/O core) are produced if the red giant branch evolution
is terminated by binary interaction before the helium core exceeds the
minimum mass for the onset of helium burning. Recently, \citet{cogr98}
found 3 faint UV-bright stars in NGC\,6397 which they suggest could be
helium-core white dwarfs \citep[supported by][]{edgr99}. Massive white
dwarfs on the other hand may evolve from blue stragglers or result from
collisions of white dwarf-binaries with subsequent merging
\citep[e.g.,][]{madh95}. \citet{saca01} discuss the effects of atmospheric 
composition and mass on the white dwarf distance determination of globular 
clusters in more detail.

Due to the faintness of these stars their study by spectroscopic
observations is still in its infancy, but first spectroscopic observations
of the white dwarf candidates in NGC~6397 \citep{mohe00}, NGC~6752, and M~4
\citep{mohe01} showed that all of them are hydrogen-rich DA white dwarfs.
Follow-up spectroscopy at better S/N should allow to derive atmospheric
parameters and thereby to verify the distances to these globular clusters. 

\section{Summary}
This section provides a brief summary of the most important points 
discussed in this paper:
\begin{itemize}
\item Abundances and rotational velocities of HB stars show a sharp change at 
temperatures of about 11,000~K to 12,000~K, with the cooler stars 
displaying
the expected cluster abundances and relatively high rotational velocities. 
The hotter stars show rather low rotational velocities
and abundances best explained by diffusion.
\item Diffusion, esp. radiative levitation of heavy elements, can most probably 
solve the problem of the low gravities found previously for HB stars 
between $\approx$11,500~K and $\approx$20,500~K.
\item The faint gap along the blue tail at $M_V\approx$\magn{3}
separates hot HB from extreme HB stars. The brighter gap at 
$M_V\approx$\magpt{0}{6} to $\approx$\magpt{1}{4} is probably caused by 
the onset of radiative levitation in
the atmospheres of the HB stars. Non-canonical evolutionary scenarios are 
probably not necessary to explain these gaps or the results of spectroscopic
analyses of hot HB/blue tail stars. 
\item The physical parameters of cool blue HB stars in metal-poor globular 
clusters agree with canonical evolutionary tracks, but yield canonical 
masses preferably for the long distance scale.
\item Hot HB stars in metal-rich globular clusters form a rather 
inhomogeneous group, that cannot be explained by {\em one} evolutionary 
scenario.
\item Hot UV bright stars selected by far-UV observations show the 
theoretically expected distribution of evolutionary stages, contrary to 
optically selected hot UV bright stars, which are biased towards luminous 
post-AGB stars. The considerable percentage of stars avoiding the thermally 
pulsing AGB might explain the lack of planetary nebulae in globular 
clusters.
\item White dwarfs in globular clusters so far have been verified to be 
hydrogen-rich DA white dwarfs.
\end{itemize}

\acknowledgements
I want to thank W.V. Dixon, W.B. Landsman, R.C. Peterson, and A.V. 
Sweigart for their careful reading of the manuscript. 
Thanks go also to R.P. Kraft for his 
suggestion to prepare this review from my habilitation thesis
and to U. Heber for his support for the publication.
I gratefully acknowledge financial support by the DLR (grant 50~OR~96029-ZA).


\begin{thebibliography}{}
\vspace*{-1.5ex}\bibitem[Altner \& Matilsky (1993)]{alma93}
Altner B., Matilsky T.A., 1993, ApJ 410, 116
\vspace*{-1.5ex}\bibitem[Arp (1955)]{arp55} 
Arp H.C., 1955, AJ 60, 317
\vspace*{-1.5ex}\bibitem[Bailyn (1995)]{bail95}
Bailyn, C., 1995, ARAA 33, 133
\vspace*{-1.5ex}\bibitem[Bailyn et al.\ (1992)]{baea92}
Bailyn C.D., Sarajedini A., Cohn, H., Lugger P., Grindlay J.E., 1992, AJ
103, 1564
\vspace*{-1.5ex}\bibitem[Barnard (1900)]{barn00} 
Barnard E.E., 1900, ApJ 12, 176
\vspace*{-1.5ex}\bibitem[Bedin et al.\ (2000)]{bepi00}
Bedin L.R., Piotto G., Zoccali M., Stetson P.B., Saviane I., Cassisi S.,
Bono G., 2000, A\&A 363, 159
\vspace*{-1.5ex}\bibitem[Behr et al.\ (1999)]{beco99}
Behr B.B., Cohen J.G., McCarthy J.K., Djorgovski S.G., 1999a, ApJ 517, L135
\vspace*{-1.5ex}\bibitem[Behr et al.\ (2000a)]{bedj99}
Behr B.B., Djorgovski S.G., Cohen J.G., et al., 2000a, ApJ 528, 849
\vspace*{-1.5ex}\bibitem[Behr et al.\ (2000b)]{beco00}
Behr B.B., Cohen J.G., McCarthy J.K., 2000b, ApJ 531, L37
\vspace*{-1.5ex}\bibitem[Bergeron et al.\ (1995a)]{bewe95}
Bergeron P., Wesemael F., Beauchamp A., 1995a, PASP 107, 1047
\vspace*{-1.5ex}\bibitem[Bergeron et al.\ (1995b)]{bewe95b}
Bergeron P., Wesemael F., Lamontagne R., Fontaine G., Saffer R.A., Allard 
N.F., 1995b, ApJ 449, 258
\vspace*{-1.5ex}\bibitem[Brocato et al.\ (1999)]{brca99}
Brocato E., Castellani V., Raimondo G., Walker A., 1999, ApJ 527, 230
\vspace*{-1.5ex}\bibitem[Brown et al.\ (1997)]{brfe97} 
Brown T.M., Ferguson H.C., Davidsen A.F., Dorman B., 1997, ApJ 482, 685
\vspace*{-1.5ex}\bibitem[Brown et al.\ (2000)]{brbo00} 
Brown T.M., Bowers C.W., Kimble R.A., Sweigart A.V., Ferguson H.C., 2000,
ApJ 532, 308
\vspace*{-1.5ex}\bibitem[Buonanno et al.\ (1985)]{buco85}
Buonanno R., Corsi C.E., Fusi Pecci F., 1985, A\&A 145, 97
\vspace*{-1.5ex}\bibitem[Buonanno et al.\ (1986)]{buca86} 
Buonanno R., Caloi V., Castellani V., Corsi C., Fusi Pecci F., Gratton R., 
 1986, A\&AS 66, 79 
\vspace*{-1.5ex}\bibitem[Buonanno et al.\ (1994)]{buco94} 
Buonanno R., Corsi C.E., Buzzoni A., Cacciari C., Ferraro F.R., Fusi Pecci 
 F., 1994, A\&A 290, 69%
\vspace*{-1.5ex}\bibitem[Buonanno et al.\ (1997)]{buco97}
Buonanno R., Corsi C.E., Bellazzini M., Ferraro F.R., Fusi Pecci F., 1997,
    AJ 113, 706
\vspace*{-1.5ex}\bibitem[Burstein et al.\ (1988)]{bube88}
Burstein D., Bertola F., Buson L.M., Faber S.M., Lauer T.R., 1988, ApJ 328, 
  440
\vspace*{-1.5ex}\bibitem[Cacciari et al.\ (1995)]{cafu95} 
Cacciari C., Fusi Pecci F., Bragaglia A., Buzzoni A., 1995, A\&A 301, 684
\vspace*{-1.5ex}\bibitem[Caloi (1972)]{calo72} 
Caloi V., 1972, A\&A 20, 357
\vspace*{-1.5ex}\bibitem[Caloi (1999)]{calo99} 
Caloi V., 1999, A\&A 343, 904
\vspace*{-1.5ex}\bibitem[Caloi (2001)]{calo01} 
Caloi V., 2001, A\&A 366, 91
\vspace*{-1.5ex}\bibitem[Carretta et al.\ (2000)]{cagr00}
Carretta E., Gratton R., Clementini G., Fusi Pecci F., 2000, ApJ 533, 215
\vspace*{-1.5ex}\bibitem[Catelan \& de Freitas Pacheco (1996)]{cadf96}
Catelan M., de Freitas Pacheco J.A., 1996, PASP 108, 166
\vspace*{-1.5ex}\bibitem[Catelan et al.\ (1998)]{cabo98} 
Catelan M., Borissova J., Sweigart A.V., Spassova N., 1998, ApJ 494, 265
\vspace*{-1.5ex}\bibitem[Charbonneau \& Michaud (1988)]{chmi88}
Charbonneau P., Michaud G., 1988, ApJ 327, 809%
\vspace*{-1.5ex}\bibitem[Charbonnel et al.\ (2000)]{chde00}
Charbonnel C., Denissenkov P.A., Weiss A., 2000, in
{\it The Galactic Halo: From Globular Clusters to Field Stars},
Proceedings of the 35$^{th}$ Li\`ege International Astrophysical Colloquium,
eds. A. Noels, P. Magain, D. Caro, E. Jehin, G. Parmentier, A. Thoul,
p. 453
\vspace*{-1.5ex}\bibitem[Code \& Welch (1979)]{cowe79} 
Code A.D., Welch G.A., 1979, ApJ 228, 95
\vspace*{-1.5ex}\bibitem[Cohen \& Gillett (1989)]{cogi89}
Cohen J.G., Gillett F.C., 1989, ApJ 346, 803
\vspace*{-1.5ex}\bibitem[Cohen \& McCarthy (1997)]{comc97}
Cohen J.G., McCarthy J.K., 1997, AJ 113, 1353
\vspace*{-1.5ex}\bibitem[Conlon et al.\ (1994)]{codu94} 
Conlon E.S., Dufton, P.L., Keenan, F.P., 1994, A\&A 290, 897
\vspace*{-1.5ex}\bibitem[Cool et al.\ (1996)]{copi96} 
Cool A.M., Piotto G., King I.R., 1996, ApJ 468, 655
\vspace*{-1.5ex}\bibitem[Cool et al.\ (1998)]{cogr98}
Cool A.M., Grindlay J.E., Cohn H.N., Lugger P.M., Bailyn C.D., 1998, ApJ
508, L75
\vspace*{-1.5ex}\bibitem[Crocker (1991)]{croc91}
Crocker D.A., 1991, in {\it The Formation and Evolution of Star Clusters},
    ed. K. Janes, ASP Conf. Ser. 13 (San Francisco), p. 253
\vspace*{-1.5ex}\bibitem[Crocker et al.\ (1988)]{crro88} 
Crocker D.A., Rood R.T., O'Connell R.W., 1988, ApJ 332, 236
\vspace*{-1.5ex}\bibitem[D'Cruz et al.\ (1996)]{dcdo96} 
D'Cruz N.L., Dorman B., Rood R.T., O'Connell R.W., 1996, ApJ 466, 359
\vspace*{-1.5ex}\bibitem[D'Cruz et al.\ (2000)]{dcoc00} 
D'Cruz N.L., O'Connell R.W., Rood R.T., et al., 2000, ApJ 530, 352
\vspace*{-1.5ex}\bibitem[de Boer (1982)]{debo82} 
de Boer K.S., 1982, A\&AS 50, 247
\vspace*{-1.5ex}\bibitem[de Boer (1985)]{debo85} 
de Boer K.S., 1985, A\&A 142, 321
\vspace*{-1.5ex}\bibitem[de Boer (1987)]{debo87} 
de Boer K.S., 1987, in {\it The 2$^{nd}$ Conference on Faint Blue Stars},
    eds. A.G.D. Philip, D.S. Hayes, J. Liebert, (L. Davis Press, 
    Schenectady), p.~95
\vspace*{-1.5ex}\bibitem[de Boer et al.\ (1995)]{dbsc95} 
de Boer K.S., Schmidt J.H.K., Heber U., 1995, A\&A 303, 95 
\vspace*{-1.5ex}\bibitem[de Boer et al.\ (1997)]{detu97}
de Boer K.S., Tucholke H.-J., Schmidt J.H.K., 1997, A\&A 317, L23
\vspace*{-1.5ex}\bibitem[Dixon et al.\ (1994)]{dida94}
Dixon W.V., Davidsen A.F., Ferguson H.C., 1994, AJ 107, 1388
\vspace*{-1.5ex}\bibitem[Dixon et al.\ (1995)]{dida95}
Dixon W.V., Davidsen A.F., Ferguson H.C., 1995, ApJ 454, L47
\vspace*{-1.5ex}\bibitem[Dixon et al.\ (1996)]{dida96} 
Dixon W.V., Davidsen A.F., Dorman B., Ferguson H.C., 1996, AJ 111, 1936
\vspace*{-1.5ex}\bibitem[Dorman (1997)]{dorm97}
Dorman B., 1997, in {\it The Nature of Elliptical Galaxies (2nd Stromlo
    Symposium)}, eds. M. Arnaboldi; G. S. Da Costa, P. Saha, ASP Conf. Ser. 
    116 (San Francisco), p. 195
\vspace*{-1.5ex}\bibitem[Dorman et al.\ (1991)]{dole91} 
Dorman, B., Lee Y.-W., VandenBerg D.A., 1991, ApJ 366, 115
\vspace*{-1.5ex}\bibitem[Dorman et al.\ (1993)]{doro93} 
Dorman B., Rood, R.T., O'Connell, W.O., 1993, ApJ 419, 596
\vspace*{-1.5ex}\bibitem[Dorman et al.\ (1995)]{dooc95}
Dorman B., O'Connell R.W., Rood R.T., 1995, ApJ 442, 105
\vspace*{-1.5ex}\bibitem[Dorman et al.\ (1997)]{dosh97} 
Dorman B., Shah R.Y., O'Connell R.W., et al., 1997, ApJ 480, L31
\vspace*{-1.5ex}\bibitem[Dreizler \& Wolff (1999)]{drwo99}
Dreizler S., Wolff B., 1999, A\&A 348, 189 
\vspace*{-1.5ex}\bibitem[Durrell \& Harris (1993)]{duha93} 
Durrell P.R., Harris W.E., 1993, AJ 105, 1420
\vspace*{-1.5ex}\bibitem[Edmonds et al.\ (1999)]{edgr99}
Edmonds P.D., Grindlay J.E., Cool A., Cohn H., Lugger P., Bailyn C., 1999, 
 ApJ 516, 250
\vspace*{-1.5ex}\bibitem[Faulkner (1966)]{faul66} 
Faulkner J., 1966, ApJ 144, 978
\vspace*{-1.5ex}\bibitem[Ferraro et al.\ (1997)]{fepa97}
Ferraro F.R., Paltrinieri B., Fusi Pecci F., Cacciari C., Dorman B., Rood R.T.,
 1997, ApJ 484, L145
\vspace*{-1.5ex}\bibitem[Ferraro et al.\ (1998)]{fepa98} 
Ferraro F.R., Paltrinieri B., Fusi Pecci F., Rood R.T., Dorman B., 1998, 
 ApJ 500, 311
\vspace*{-1.5ex}\bibitem[Fontaine \& Chayer (1997)]{foch97}
Fontaine G., Chayer P., 1997, in {\it The 3$^{rd}$ Conf. on Faint
Blue Stars}, eds. A.G.D. Philip, J. Liebert \& R.A. Saffer 
(L. Davis Press, Schenectady), p. 169
\vspace*{-1.5ex}\bibitem[Fusi Pecci et al.\ (1993)]{fufe93} 
Fusi Pecci F., Ferraro F.R., Bellazzini M., Djorgovski S., Piotto G., 
 Buonanno R., 1993, AJ 105, 1145
\vspace*{-1.5ex}\bibitem[Gilliland et al.\ (2000)]{gibr00} 
Gilliland R.L., Brown T.M., Guhathakurta P., Sarajedini A., Milone E.F.,
Albrow M.D., Baliber N.R., et al., 2000, ApJ 545, L47
\vspace*{-1.5ex}\bibitem[Gingold (1976)]{ging76} 
Gingold R.A., 1976, ApJ 204, 116
\vspace*{-1.5ex}\bibitem[Glaspey et al.\ (1985)]{glde85}
Glaspey J.W., Demers S., Moffat A.F.J., Shara M., 1985, ApJ 289, 326
\vspace*{-1.5ex}\bibitem[Glaspey et al.\ (1986)]{glde86}
Glaspey J.W., Demers S., Moffat A.F.J., Michaud G., 1986, PASP 98, 1123
\vspace*{-1.5ex}\bibitem[Glaspey et al.\ (1989)]{glmi89}
Glaspey J.W., Michaud G., Moffat A.F.J., Demers S., 1989, ApJ 339, 926
\vspace*{-1.5ex}\bibitem[Graham \& Doremus (1966)]{grdo66}
Graham J.A., Doremus C., 1966, AJ 73, 226
\vspace*{-1.5ex}\bibitem[Gratton et al.\ (2001)]{grbo01}
Gratton R.G., Bonifacio P., Bragaglia A., Carretta E., Castellani V.,
Centurion M.,Chieffi A., et al., 2001, A\&A 369, 87
\vspace*{-1.5ex}\bibitem[Greenstein (1939)]{gree39} 
Greenstein J.L., 1939, ApJ 90, 387
\vspace*{-1.5ex}\bibitem[Greenstein (1971)]{gree71} 
Greenstein J.L., 1971, in {\it White Dwarfs}, ed. W.J. Luyten,
    IAU Symp. 42, (Reidel), p. 46
\vspace*{-1.5ex}\bibitem[Greenstein \& M\"unch (1966)]{grmu66}
Greenstein G.S., M\"unch G., 1966, ApJ 146, 518
\vspace*{-1.5ex}\bibitem[Greenstein et al.\ (1967)]{grtr67}
Greenstein G.S., Truran J.W., Cameron A.G.W., 1967, Nature 213, 871
\vspace*{-1.5ex}\bibitem[Greggio \& Renzini (1990)]{grre90} 
Greggio L., Renzini A., 1990, ApJ 364, 35%
\vspace*{-1.5ex}\bibitem[Greggio \& Renzini (1999)]{grre99} 
Greggio L., Renzini A., 1999, Mem. S.A.I. 70, 691%
\vspace*{-1.5ex}\bibitem[Grundahl et al.\ (1999)]{grca99}
Grundahl F., Catelan M., Landsman W.B., Stetson P.B., Andersen M., 1999,
   ApJ 524, 242
\vspace*{-1.5ex}\bibitem[Harris et al.\ (1983)]{hane83} 
Harris H.C., Nemec J.M., Hesser J.E., 1983, PASP 95, 256
\vspace*{-1.5ex}\bibitem[Heber (1986)]{hebe86}
Heber U., 1986, A\&A 155, 33
\vspace*{-1.5ex}\bibitem[Heber (1987)]{hebe87} 
Heber U., 1987, Mitt. Astron. Ges. 70, 79
\vspace*{-1.5ex}\bibitem[Heber et al.\ (1984)]{hehu84}
Heber U., Hunger K., Jonas G., Kudritzki R.P., 1984, A\&A 130, 119
\vspace*{-1.5ex}\bibitem[Heber et al.\ (1986)]{heku86} 
Heber U., Kudritzki R.P., Caloi V., Castellani V., Danziger J., Gilmozzi R., 
 1986, A\&A 162, 171
\vspace*{-1.5ex}\bibitem[Heber \& Kudritzki (1986)]{heku86b} 
Heber U., Kudritzki R.P., 1986, A\&A 169, 244
\vspace*{-1.5ex}\bibitem[Heber et al.\ (1993)]{hedr93} 
Heber U., Dreizler S., Werner, K., 1993, Acta Astron. 43, 337
\vspace*{-1.5ex}\bibitem[Heber et al.\ (1997)]{hemo97} 
Heber U., Moehler S., Reid I.N., 1997, in {\it HIPPARCOS Venice '97},
    ed. B. Battrick, ESA-SP 402, p. 461 
\vspace*{-1.5ex}\bibitem[Hill et al.\ (1996)]{hich96}
Hill R.S., Cheng K.-P., Smith E.P., et al., 1996, AJ 112, 601
\vspace*{-1.5ex}\bibitem[Hoyle \& Schwarzschild (1955)]{hosc55} 
Hoyle F., Schwarzschild M., 1955, ApJS 2, 1
\vspace*{-1.5ex}\bibitem[Hui-Bon-Hoa et al.\ (2000)]{hule00}
Hui-Bon-Hoa A., LeBlanc F., Hauschildt P., 2000, ApJ 535, L43
\vspace*{-1.5ex}\bibitem[Iben \& Rood (1970)]{ibro70} 
Iben I.Jr., Rood R.T., 1970, ApJ 161, 587
\vspace*{-1.5ex}\bibitem[Iben \& Tutukov (1984)]{ibtu84} 
Iben I. Jr., Tutukov A.V., 1984, ApJS 54, 335
\vspace*{-1.5ex}\bibitem[Iben \& Tutukov (1993)]{ibtu93} 
Iben I. Jr., Tutukov A.V., 1993, ApJ 418, 343
\vspace*{-1.5ex}\bibitem[Jacoby et al.\ (1997)]{jamo97} 
Jacoby G.H., Morse J. A., Fullton L.K., Kwitter K.B., Henry R.B.C, 1997, 
    AJ 114, 2611
\vspace*{-1.5ex}\bibitem[Kaluzny \& Udalski (1992)]{kaud92}
Kaluzny J., Udalski A., 1992, AcA 42, 29
\vspace*{-1.5ex}\bibitem[Kaluzny et al.\ (1997)]{kakr97}
Kaluzny J., Krzemis\'nki W., Mazur B., Wysocka A., Stepie\'n K., 1997,
AcA 47, 249
\vspace*{-1.5ex}\bibitem[Kraft (1994)]{kraf94} 
Kraft R.P., 1994, PASP 106, 553 
\vspace*{-1.5ex}\bibitem[Kraft et al.\ (1997)]{krsn97} 
Kraft R.P., Sneden C., Smith G.H., Shetrone M.D., Langer G.E., Pilachowski 
 C.A., 1997, AJ 113, 279
\vspace*{-1.5ex}\bibitem[Kurucz (1992)]{kuru92} 
Kurucz R.L., 1992, in {\it The Stellar Populations of Galaxies},
    eds. B. Barbuy \& A. Renzini, IAU Symp. 149 (Kluwer:Dordrecht), 225
\vspace*{-1.5ex}\bibitem[Lambert et al.\ (1992)]{lamc92}
Lambert D.L., McWilliam A., Smith V.V., 1992, ApJ 386, 685
\vspace*{-1.5ex}\bibitem[Landsman et al.\ (1996)]{lasw96} 
Landsman W.B., Sweigart A.V., Bohlin R.C., et al., 1996, ApJ 472, L93
\vspace*{-1.5ex}\bibitem[Landsman et al.\ (1998)]{labo98}
Landsman W.B., Bohlin R.C., Neff S.G., et al., 1998, AJ 116, 789
\vspace*{-1.5ex}\bibitem[Landsman et al.\ (2001)]{wayne}
Landsman W.B., Moehler S.,Napiwotzki R., Sweigart A., Heber U., 
Catelan M., Stecher T., 2001, in prep.
\vspace*{-1.5ex}\bibitem[Layden et al.\ (1999)]{lari99}
Layden A.C., Ritter L.A., Welch D.L., Webb T.M.A., 1999, AJ 117, 1313
\vspace*{-1.5ex}\bibitem[Lee et al.\ (1994)]{lede94} 
Lee Y.-W., Demarque P., Zinn R., 1994, ApJ 423, 248
\vspace*{-1.5ex}\bibitem[Liebert et al.\ (1994)]{lisa94} 
Liebert J., Saffer R.A., Green E.M., 1994, AJ 107, 1408
\vspace*{-1.5ex}\bibitem[Markov et al.\ (2001)]{masp01}
Markov H.S., Spassova N.M., Baev P.V., 2001, MNRAS in press 
({\tt astro-ph/0103245})
\vspace*{-1.5ex}\bibitem[Marsh et al.\ (1995)]{madh95}
Marsh T.R., Dhillon V.S., Duck S.R., 1995, MNRAS 275, 828
\vspace*{-1.5ex}\bibitem[Michaud (1982)]{mich82}
Michaud G., 1982, ApJ 258, 349
\vspace*{-1.5ex}\bibitem[Michaud et al.\ (1983)]{miva83}
Michaud G., Vauclair G., Vauclair S., 1983, ApJ 267, 256
\vspace*{-1.5ex}\bibitem[Michaud et al.\ (1989)]{mibe89}
Michaud G., Bergeron P., Heber U., Wesemael F., 1989, ApJ 338, 417
\vspace*{-1.5ex}\bibitem[Mihalas (1966)]{miha66}
Mihalas D.M., 1966, ApJS 13, 1
\vspace*{-1.5ex}\bibitem[Moehler (1999)]{moeh99}
Moehler S., 1999, Reviews in Modern Astronomy, ed. R. Schielicke, Vol. 12, 
p.~281
\vspace*{-1.5ex}\bibitem[Moehler et al.\ (1990b)]{mohe90}
Moehler S., Heber U., de Boer K.S., 1990b, A\&A 239, 265
\vspace*{-1.5ex}\bibitem[Moehler et al.\ (1994)]{mohe94}
Moehler S., Heber U., de Boer K.S., 1994, in {\it Hot Stars in the Halo},
    eds. S.J. Adelman, A. Upgren \& C.J. Adelman, (Cambridge University Press,
    Cambridge), p. 217
\vspace*{-1.5ex}\bibitem[Moehler et al.\ (1995)]{mohe95} 
Moehler S., Heber U., de Boer K.S., 1995, A\&A 294, 65 
\vspace*{-1.5ex}\bibitem[Moehler et al.\ (1997a)]{mohe97a} 
Moehler S., Heber U., Durrell P., 1997a, A\&A 317, L83
\vspace*{-1.5ex}\bibitem[Moehler et al.\ (1997b)]{mohe97b} 
Moehler S., Heber U., Rupprecht G., 1997b, A\&A 319, 109
\vspace*{-1.5ex}\bibitem[Moehler et al.\ (1998a)]{mola98} 
Moehler S., Landsman W., Napiwotzki R., 1998a, A\&A 335, 510
\vspace*{-1.5ex}\bibitem[Moehler et al.\ (1998b)]{mohe98b} 
Moehler S., Heber U., Lemke M., Napiwotzki R., 1998b, A\&A 339, 537
\vspace*{-1.5ex}\bibitem[Moehler et al.\ (1999)]{mosw99}
Moehler S., Sweigart A.V., Catelan M., 1999, A\&A 351, 519
\vspace*{-1.5ex}\bibitem[Moehler et al.\ (2000a)]{mohe00}
Moehler S., Heber U., Napiwotzki R., Koester D., Renzini A., 2000a, A\&A
354, L75
\vspace*{-1.5ex}\bibitem[Moehler et al.\ (2000b)]{mosw00}
Moehler S., Sweigart A.V., Landsman W., Heber U., 2000b, A\&A 360, 120
\vspace*{-1.5ex}\bibitem[Moehler et al.\ (2000c)]{mola00}
Moehler S., Landsman W., Dorman B., 2000c, A\&A 361, 937
\vspace*{-1.5ex}\bibitem[Moehler et al.\ (2001)]{mohe01}
Moehler S., Heber U., Napiwotzki R., Koester D., Renzini A., 2001, in prep.
\vspace*{-1.5ex}\bibitem[Newell (1973)]{newe73} 
Newell E.B., 1973, ApJS 26, 37
\vspace*{-1.5ex}\bibitem[Newell \& Graham (1976)]{negr76}
Newell E.B., Graham J.A., 1976, ApJ 204, 804
\vspace*{-1.5ex}\bibitem[O'Connell et al.\ (1997)]{ocdo97} 
O'Connell R.W., Dorman B., Shah R.Y., et al., 1997, AJ 114, 1982
\vspace*{-1.5ex}\bibitem[Paresce et al.\ (1995)]{pade95}
Paresce F., de Marchi G., Romaniello M., 1995, ApJ 440, 216
\vspace*{-1.5ex}\bibitem[Parise et al.\ (1998)]{pabo98} 
Parise R.A., Bohlin R.C., Neff S.G., et al., 1998, ApJ 501, L67
\vspace*{-1.5ex}\bibitem[Pease (1928)]{peas28} 
Pease F.G., 1928, PASP 40, 342
\vspace*{-1.5ex}\bibitem[Peterson (1983)]{pete83}
Peterson R.C., 1983, ApJ 275, 737
\vspace*{-1.5ex}\bibitem[Peterson (1985a)]{pete85a}
Peterson R.C., 1985a, ApJ 289, 320
\vspace*{-1.5ex}\bibitem[Peterson (1985b)]{pete85b}
Peterson R.C., 1985b, ApJ 294, L35
\vspace*{-1.5ex}\bibitem[Peterson et al.\ (1995)]{pero95} 
Peterson R.C., Rood R.T., Crocker D.A., 1995, ApJ 453, 214%
\vspace*{-1.5ex}\bibitem[Peterson et al.\ (2000)]{pero00}
Peterson R.C., Rood R.T., Crocker D.A., Kraft R.P., 2000, in
{\it The Galactic Halo: From Globular Clusters to Field Stars},
Proceedings of the 35$^{th}$ Li\`ege International Astrophysical Colloquium,
eds. A. Noels, P. Magain, D. Caro, E. Jehin, G. Parmentier, A. Thoul, p. 523
\vspace*{-1.5ex}\bibitem[Piotto et al.\ (1997)]{piot97}
Piotto G., Sosin C., King I.R., et al., 1997, in {\it Advances in Stellar
    Evolution}, eds. Rood R.T., Renzini A., (Cambridge University Press, 
    Cambridge), p.~84
\vspace*{-1.5ex}\bibitem[Piotto et al.\ (1999)]{pizo99}
Piotto G., Zoccali M., King I.R., Djorgovski S.G., Sosin C.,
Rich R.M., Meylan G., 1999, AJ 118, 1727
\vspace*{-1.5ex}\bibitem[Pritzl et al.\ (1999)]{prsm99}
Pritzl B., Smith H.A., Catelan M., Sweigart A.V., 2000, ApJ 530, L41
\vspace*{-1.5ex}\bibitem[Reid (1999)]{reid99} 
Reid I.N. 1999, ARA\&A 37, 191
\vspace*{-1.5ex}\bibitem[Remillard et al.\ (1980)]{reca80}
Remillard R.A., Canizares C.R., McClintock J.E., 1980, ApJ 240, 109
\vspace*{-1.5ex}\bibitem[Renzini (1985)]{renz85}
Renzini A., 1985, in {\it Horizontal Branch and UV-Bright Stars}, ed. 
    A.G. Davis Philip (L. Davis Press, Schenectady), p. 19
\vspace*{-1.5ex}\bibitem[Renzini et al.\ (1996)]{rebr96} 
Renzini A., Bragaglia A., Ferraro F.R., et al., 1996, ApJ 465, L23
\vspace*{-1.5ex}\bibitem[Rich et al.\ (1993)]{rimi93}
Rich R.M., Minniti D., Liebert J., 1993, ApJ 406, 489
\vspace*{-1.5ex}\bibitem[Rich et al.\ (1997)]{riso97} 
Rich R.M., Sosin C., Djorgovski S.G., et al., 1997, ApJ 484, L25
\vspace*{-1.5ex}\bibitem[Richer et al.\ (1995)]{rifa95} 
Richer H.B., Fahlmann G.G., Ibata R.A., et al., 1995, ApJ 451, L17
\vspace*{-1.5ex}\bibitem[Richer et al.\ (1997)]{rifa97} 
Richer H.B., Fahlmann G.G., Ibata R.A., et al., 1997, ApJ 484, 741
\vspace*{-1.5ex}\bibitem[Rood (1973)]{rood73} 
Rood R.T., 1973, ApJ 184, 815
\vspace*{-1.5ex}\bibitem[Rood \& Crocker (1989)]{rocr89}
 Rood R.T., Crocker D.A., 1989, in {\it The Use of Pulsating Stars in 
    Fundamental Problems of Astronomy}, ed. E.G. Schmidt (Cambridge University 
    Press, Cambridge), p.~103
\vspace*{-1.5ex}\bibitem[Rood et al.\ (1997)]{rowh97}
Rood R.T., Whitney J., D'Cruz N., 1997, in {\it Advances in Stellar
 Evolution}, eds. R.T. Rood \& A. Renzini (Cambridge: CUP), p. 74 
\vspace*{-1.5ex}\bibitem[Rose \& Deng (1999)]{rode99}
Rose J.A., Deng S., 1999, AJ 117, 2213
\vspace*{-1.5ex}\bibitem[Saffer et al.\ (1994)]{sabe94} 
Saffer R.A., Bergeron P., Koester D., Liebert J., 1994, ApJ 432, 351
\vspace*{-1.5ex}\bibitem[Saffer et al.\ (1997)]{sake97}
Saffer R.A., Keenan F.P., Hambly N.C., Dufton P.L., Liebert J., 1997, ApJ 491,
172
\vspace*{-1.5ex}\bibitem[Salaris et al.\ (2001)]{saca01} 
Salaris M., Cassisi S., Garcia-Berro E., Isern J., Torres S., 2001,
A\&A in press ({\tt astro-ph/0103315})
\vspace*{-1.5ex}\bibitem[Sandage \& Wallerstein (1960)]{sawa60} 
Sandage A.R., Wallerstein G., 1960, ApJ 131, 598%
\vspace*{-1.5ex}\bibitem[Sargent (1967)]{sarg67}
Sargent W.L.W., 1967, ApJ 148, L147
\vspace*{-1.5ex}\bibitem[Sch\"onberner (1983)]{scho83} 
Sch\"onberner D., 1983, ApJ 272, 708
\vspace*{-1.5ex}\bibitem[Schwarzschild \& H\"arm (1970)]{scha70} 
Schwarzschild M., H\"arm R., 1970, ApJ 160, 341
\vspace*{-1.5ex}\bibitem[Searle \& Rodgers (1966)]{sero66}
Searle L., Rodgers A.W., 1966, ApJ 143, 809
\vspace*{-1.5ex}\bibitem[Shapley (1915a)]{shap15a} 
Shapley H., 1915a, Contr. Mt. Wilson 115
\vspace*{-1.5ex}\bibitem[Shapley (1915b)]{shap15b} 
Shapley H., 1915b, Contr. Mt. Wilson 116
\vspace*{-1.5ex}\bibitem[Shapley (1930)]{shap30} 
Shapley H., 1930, {\it Star Clusters}, (The Maple Press Company, York, 
    Pennsylvania, USA)
\vspace*{-1.5ex}\bibitem[Sills \& Pinsonneault (2000)]{sipi00}
Sills A., Pinsonneault M.H., 2000, ApJ 540, 489
\vspace*{-1.5ex}\bibitem[Soker (1998)]{soke98}
Soker N., 1998, AJ 116, 1308
\vspace*{-1.5ex}\bibitem[Soker \& Harpaz (2000)]{soha00}
Soker N., Harpaz A., 2000, MNRAS 317, 861
\vspace*{-1.5ex}\bibitem[Sosin et al.\ (1997)]{sodo97}
Sosin C., Dorman B., Djorgovski S.G., et al., 1997, ApJ 480, L35
\vspace*{-1.5ex}\bibitem[Stecher et al.\ (1997)]{stco97} 
Stecher T., Cornett R.H., Greason M.R., et al., 1997, PASP 109, 584
\vspace*{-1.5ex}\bibitem[Stoeckley \& Greenstein (1968)]{stgr68} 
Stoeckley R., Greenstein J.L., 1968, ApJ 154, 909%
\vspace*{-1.5ex}\bibitem[Storm et al.\ (1994)]{stca94}
Storm J., Carney B.W., Latham D.W., 1994, A\&A 290, 443
\vspace*{-1.5ex}\bibitem[Strom \& Strom (1970)]{stst70} 
Strom S.E., Strom K.M., 1970, ApJ 159, 195
\vspace*{-1.5ex}\bibitem[Strom et al.\ (1970)]{stst70b} 
Strom S.E., Strom K.M., Rood R.T., Iben I.Jr., 1970, A\&A 8, 243
\vspace*{-1.5ex}\bibitem[Sweigart (1987)]{swei87} 
Sweigart A.V., 1987, ApJS 65, 95
\vspace*{-1.5ex}\bibitem[Sweigart (1994)]{swei94} 
Sweigart A.V., 1994, in {\it Hot Stars in the Galactic Halo}, eds. S.
Adelman, A. Upgren, C.J. Adelman, (Cambridge University Press, 
Cambridge), p. 17
\vspace*{-1.5ex}\bibitem[Sweigart (1997a)]{swei97a} 
Sweigart A.V., 1997a, ApJ 474, L23
\vspace*{-1.5ex}\bibitem[Sweigart (1997b)]{swei97b} 
Sweigart A.V. 1997b, in {\it The 3$^{rd}$ Conf. on Faint
Blue Stars}, eds. A.G.D. Philip, J. Liebert \& R.A. Saffer 
(L. Davis Press, Schenectady), p. 3%
\vspace*{-1.5ex}\bibitem[Sweigart (2001)]{swei01} 
Sweigart A.V., 2001, to appear in Highlights of Astronomy, Vol. 12 
({\tt astro-ph/0103133})
\vspace*{-1.5ex}\bibitem[Sweigart et al.\ (1974)]{swme74} 
Sweigart A.V., Mengel J.G., Demarque P., 1974, A\&A 30, 13
\vspace*{-1.5ex}\bibitem[Sweigart \& Gross (1974)]{swgr74} 
Sweigart A.V., Gross P.G., 1974, ApJ, 190, 101
\vspace*{-1.5ex}\bibitem[Sweigart \& Gross (1976)]{swgr76} 
Sweigart A.V., Gross P.G., 1976, ApJS 32, 367
\vspace*{-1.5ex}\bibitem[Sweigart \& Catelan (1998)]{swca98} 
Sweigart A.V., Catelan M., 1998, ApJ 501, L63
\vspace*{-1.5ex}\bibitem[ten Bruggencate (1927)]{tebr27} 
ten Bruggencate P., 1927, {\it Sternhaufen}, (Julius Springer Verlag,
    Berlin)
\vspace*{-1.5ex}\bibitem[Testa et al.\ (2001)]{teco01}
Testa V., Corsi C.E., Andreuzzi G.,  et al., 2001, AJ 121, 916
\vspace*{-1.5ex}\bibitem[Traving (1962)]{trav62} 
Traving G., 1962, ApJ 135, 439
\vspace*{-1.5ex}\bibitem[Tuchman (1985)]{tuch85} 
Tuchman Y., 1985, ApJ 288, 248
\vspace*{-1.5ex}\bibitem[Unglaub \& Bues (1998)]{unbu98}
Unglaub K., Bues I., 1998, A\&A 338, 75
\vspace*{-1.5ex}\bibitem[Vink et al.\ (1999)]{vihe99}
Vink J.S., Heap S.R., Sweigart A.V., Lanz T., Hubeny I., 1999, A\&A 345, 109
\vspace*{-1.5ex}\bibitem[Walker (1999)]{walk99}
Walker A.R., 1999, AJ 118, 432 (Erratum: AJ 119, 1512)
\vspace*{-1.5ex}\bibitem[Whitney et al.\ (1998)]{whro98} 
Whitney J.H., Rood R.T., O'Connell R.W., et al., 1998, ApJ 495, 284
\vspace*{-1.5ex}\bibitem[Yong et al.\ (2000)]{yode00}
Yong H., Demarque P., Yi S., 2000, ApJ 539, 928
\vspace*{-1.5ex}\bibitem[Zinn (1974)]{zinn74} 
Zinn R., 1974, ApJ, 193, 593
\vspace*{-1.5ex}\bibitem[Zinn et al.\ (1972)]{zine72} 
Zinn R.J., Newell E.B., Gibson J.B., 1972, A\&A 18, 390
\vspace*{-1.5ex}\bibitem[Zoccali et al.\ (2001)]{zore01}
Zoccali M., Renzini A., Ortolani S., Bragaglia A., Bohlin R., Carretta E.,
et al., 2001, ApJ in press ({\tt astro-ph/0101485})
\end{thebibliography}
\end{document}